\documentclass[aps,prb,reprint]{revtex4-2}

\usepackage{amsmath,amsfonts,amssymb,graphicx,braket,bbold,mathtools,array}
\usepackage{float}
\usepackage{multirow}
\usepackage[]{units}
\usepackage{hyperref}
\usepackage{MnSymbol}
\hypersetup{
    colorlinks=true,
    linkcolor=blue,
    filecolor=blue,      
    urlcolor=blue,
    citecolor=blue
}

\usepackage[table]{xcolor}

\usepackage[T1]{fontenc}
\usepackage{tikz}
\usetikzlibrary{calc,patterns,decorations.pathmorphing,decorations.markings, shapes, positioning, shapes.geometric}

\newcommand*{\figref}[2][]{%
  \hyperref[{fig:#2}]{%
   \ref*{fig:#2}%
    \ifx\\#1\\%
    \else #1%
    \fi
  }%
}

\begin{document}
\title{Theory of qubit noise characterization using the long-time cavity transmission}
\author{Philipp M. Mutter}
\email{philipp.mutter@uni-konstanz.de}
\author{Guido Burkard}
\email{guido.burkard@uni-konstanz.de}
\affiliation{Department of Physics, University of Konstanz, D-78457 Konstanz, Germany}

\begin{abstract}
Noise induced decoherence is one of the main threats to large-scale quantum computation. In an attempt to assess the noise affecting a qubit we go beyond the standard steady-state solution of the transmission through a qubit-coupled cavity in input-output theory by including dynamical noise in the description of the system. We solve the quantum Langevin equations exactly for a noise-free system and treat the noise as a perturbation. In the long-time limit the corrections may be written as a sum of convolutions of the noise power spectral density with an integration kernel that depends on external control parameters. Using the convolution theorem, we invert the corrections and obtain relations for the noise spectral density as an integral over measurable quantities. Additionally, we treat the noise exactly in the dispersive regime, and again find that noise characteristics are imprinted in the long-time transmission in convolutions containing the power spectral density.
\end{abstract}

\maketitle

\section{Introduction}
In the absence of quantum error correction~\cite{Preskill2018}, fluctuations from the environment severely impair the performance of quantum hardware, as such external noise leads to decoherence, i.e., the permanent loss of quantum coherence~\cite{Zurek2003}. Even with quantum error correction at hand, the presence of noise can force the system to exceed the critical error threshold of fault tolerant quantum computation~\cite{Shor1996,Knill1998,Kitaev2003,Dawson2006, Aharonov2008}, thereby rendering large-scale computations impossible~\cite{Nielsen2010}.  To mitigate the effects of the environment, it can be advantageous to have a precise knowledge of the form of the temporal fluctuations. As a consequence, the determination of noise characteristics is an important task in present-day quantum information processing units.

All qubit realizations suffer from noise in one way or the other, either directly or indirectly. This holds especially for solid-state qubits which are surrounded by a macroscopic number of atoms that make up the host crystal~\cite{Dutta1981review,Chirolli2008,Bergli2009,Paladino2014review}. Semiconductor charge qubits in quantum dots suffer from charge noise stemming from fluctuating gate voltages~\cite{Ithier2005,Russ2015,Kranz2020}, spin qubits suffer from effective random magnetic fields due to the interaction with the host atomic nuclei~\cite{Tyryshkin2006,Bluhm2011,Kuhlmann2013,Malinowski2017}, and superconducting circuits can be affected by magnetic flux noise, quasiparticles and two-level fluctuators~\cite{Clarke2008,Oliver2013,Krantz2019,deGraaf2020}. Via the spin-orbit interaction, charge noise can affect the spin degree of freedom as well~\cite{Huang2014, Bermeister2014,Connors2022}. Hybrid systems such as flopping mode qubits or other spin qubits that rely on the spin-orbit interaction for gate operations, e.g., hole systems in germanium~\cite{Hendrickx2020b,Mutter2020cavitycontrol, Mutter2021natural, Jirovec2021, Mutter2021_ST_qubit, Jirovec2022}, may even suffer from both magnetic and charge noise~\cite{Benito2019b}. In general, enhanced qubit performance from an extended control parameter space typically comes at the price of additional decoherence channels and increased susceptibility of the logical two-level system to noise.

\begin{figure}
	\includegraphics[scale=0.24]{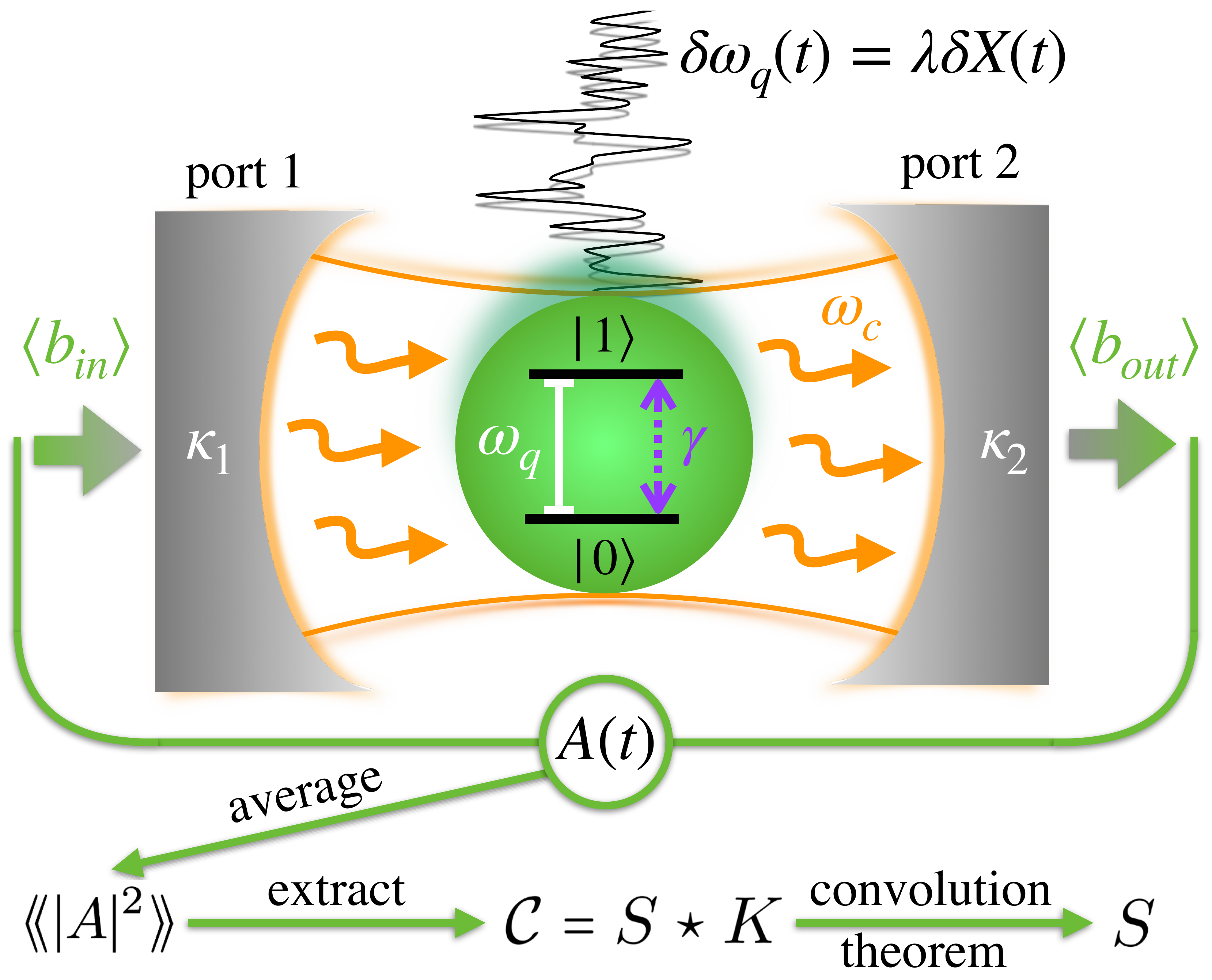}
	\caption{The system under consideration consists of a cavity-coupled qubit (green dot) with fluctuating energy separation $\omega_q$ due to noise $\delta X(t)$ affecting its control parameters. The cavity mirrors (grey) allow for an input field $b_{\rm in}$ at the left port and an output field $b_{\rm out}$ at the right port, thereby creating a measurable transmission signal $A(t)$. We find that the averaged transmission probability $\llangle \vert A \vert^2 \rrangle$ receives noise corrections in the form of sums over convolutions $\mathcal{C} =S \star K$ of the power spectral density $S(\omega)$ of $\delta X$ with an integration kernel $K$. As a result, the spectral density can be extracted from the averaged transmission probability by using the convolution theorem.}
	\label{fig:system_schematic}
\end{figure}

Here, we address the problem of characterizing fluctuations in a quantum bit (qubit), which represents the fundamental building block of a quantum computer.  To obtain information about the noise power spectral density, we propose a scheme based on the long-time transmission through an electromagnetic cavity interacting with the qubit (Fig.~\ref{fig:system_schematic}). The steady-state transmission through a resonator-qubit system has become an important tool in extracting system parameters such as valley splittings~\cite{Burkard2016,Mi2017} and in qubit readout~\cite{Blais2004,Schuster2005, Schuster2007,Bianchetti2009,DAnjou2019,Mielke2021}. The cavity transmission signal has played an important role in providing evidence for the achievement of the strong coupling regime of cavity quantum electrodynamics for superconducting coplanar microwave cavities coupled to either superconducting qubits~\cite{Wallraff2004} or semiconducting qubits \cite{Viennot2015, Stockklauser2017, Mi2018, Koski2020} and has been used to experimentally characterize charge noise in a semiconductor double quantum dot~\cite{Basset2014}. Recent studies have theoretically investigated the effect of general dynamical noise on the transient cavity transmission in the dispersive regime~\cite{Mutter2022}, as well as the effect of quantum noise~\cite{McIntyre2022arXiv}. Here, we study the effect of dynamical noise on the long-time transmission. We complement the analysis in Ref.~\cite{Mutter2022} in the dispersive regime, and in addition obtain results that are valid for arbitrary qubit-photon detunings given that the decay rates in the system are sufficiently large to allow for a quickly converging perturbation series in the dynamical noise parameter. In both regimes we find that the averaged transmission probability contains measurable information on spectral features of the noise in the system.

The remainder of the paper is structured as follows: Sec.~\ref{sec:model} introduces the theoretical model and displays the quantum Langevin equations that we aim to solve in different regimes in the subsequent sections. We treat the noise as a perturbation in Sec.~\ref{sec:perturbation_theory_in_noise} and distinguish between two cases: In Sec.~\ref{sec:diagonalizable_L} we work in the regime where the matrix $L$ appearing in the system of differential equations is diagonalizable, while we focus on the special parameter regime in which $L$ becomes non-diagonalizable in Sec.~\ref{sec:non-diagonalizable_L}.  In both cases we find that the long-time transmission receives noise corrections which can be expressed as convolutions of the spectral density with integration kernels depending on system parameters. We discuss and compare our findings in Sec.~\ref{subsec:discussion}, and show how the noise power spectral density can be extracted in real measurements by using the convolution theorem in Sec.~\ref{sec:extracting_noise_characteristics}. Additionally, we investigate the long-time limit in the dispersive regime in Sec.~\ref{sec:perturbation_theory_in_g} and thereby complement the results on the transient transmission in Ref.~\cite{Mutter2022}. Again, we find that the cavity transmission contains information on the fluctuations affecting the qubit in terms of convolutions containing the noise power spectral density. Finally, Sec.~\ref{sec:conclusions} provides a conclusion and an outlook on possible future research directions.

\section{System and Transmission}
\label{sec:model}

We study a generic noisy two-level system with fluctuating qubit energy separation $\omega_q + \delta \omega_q(t)$ and total noise-independent intrinsic decoherence rate $\gamma = \coth (\omega_q/2T) \gamma_1/2 +  \gamma_{\varphi}$, where $T$ is the temperature, $\gamma_1$ is the relaxation rate at zero Kelvin and $\gamma_{\varphi}$ is the dephasing rate~\footnote{The intrinsic decoherence rate $\gamma$ used here is taken to be half the rate used in Ref.~\cite{Mutter2022} for consistency with the relation $\gamma \equiv T_2^{-1} = (2T_1)^{-1} + T_{\varphi}^{-1}$ in Bloch-Redfield theory.}. The fluctuations may be written in leading order as $\delta \omega_q = \lambda \delta X(t)$, where $\lambda = \partial_{X} \omega_q \vert_{\delta X = 0}$ is the noise sensitivity of the qubit, $X$ is the noise-affected qubit control parameter, and $\delta X(t)$ describes the time-dependent classical random fluctuations of this parameter. We remark that fluctuations of the coupling constant $g$ may also be present in the system but can be controlled externally (e.g. via the detuning in a charge qubit, see below) and we focus on the regime where they may be neglected. The qubit is allowed to interact with a single cavity mode of frequency $\omega_c$, and the qubit-cavity coupling is assumed to be linear with coupling strength $g$ (Fig.~\ref{fig:system_schematic}).  The cavity decay rate is given by $\kappa = \kappa_1 + \kappa_2 + \kappa_{\rm int}$, where $\kappa_j$ is the decay rate at port $j \in \lbrace 1,2 \rbrace$ and $\kappa_{\rm int}$ the intrinsic photon loss rate, and in the following we choose a classical input field of amplitude $\langle b_{\text{in}} \rangle $ and frequency $\omega_p$ to be present at port 1, while no input field is assumed to be present at port 2. In a frame corotating with the probe frequency $\omega_p$ and within the rotating wave approximation, the quantum Langevin equations for the expectation values of the spin ladder operator $\sigma_-$ and the photon annihilation operator $a$ take the form,
	\begin{align}
	\label{eq:deq_matrix_form}
	\begin{split}
	&\frac{d}{dt} \begin{pmatrix}
		\langle \sigma_- \rangle \\
		\langle a \rangle
	\end{pmatrix} 
	+\left[ L + \delta L (t)  \right] \begin{pmatrix}
		\langle \sigma_- \rangle \\
		\langle a \rangle
	\end{pmatrix} 
	= 
	\begin{pmatrix}
		0 \\
		\sqrt{\kappa_1} \langle b_{\text{in}} \rangle 
	\end{pmatrix} , \\
	& L  = \begin{pmatrix}
		i \Delta_q + \gamma/2 & -i \langle \sigma_z \rangle g \\
		i g & i \Delta_c + \kappa/2
	\end{pmatrix},
	\; \delta L (t)  = \begin{pmatrix}
		i \lambda \delta X(t) & 0 \\
		0 & 0
	\end{pmatrix},	 
	\end{split}
	\end{align}
where we introduce the probe-qubit (probe-cavity) detuning $\Delta_q = \omega_q - \omega_p$ ($\Delta_c = \omega_c - \omega_p$). The dynamics of the operators under consideration are governed by the non-Hermitian system matrix $L$ and the noise corrections appearing in the dynamical matrix $\delta L (t)$. We remark that $\langle \sigma_z \rangle \equiv \langle \sigma_z^{(0)} \rangle $ is taken to be the zeroth-order term in the photon-qubit coupling $g$ while higher orders are neglected, a standard assumption in input-output theory~\cite{Burkard2016, Benito2017_input_output, Mielke2021}, which is justified as follows: The only equation featuring $\langle \sigma_z \rangle$ is the one for $\langle \sigma_- \rangle$ in the term $g \langle \sigma_z a \rangle$ which to leading order in $g$ reads $g \langle \sigma_z^{(0)} \rangle \langle a \rangle$ with exact $\langle a \rangle$ for separable initial qubit-cavity states. As the solution of $\langle \sigma_- \rangle$ is only required to first order in $g$ in the expansion of $\langle \sigma_z \rangle$ to calculate the transmission via the exact Langevin equation for $\langle a \rangle$, the procedure is consistent. Physically this means that the level populations are assumed to be unaffected by the interaction with the photons. In the remainder of this paper we assume a constant thermal distribution of the qubit energy levels, $\langle \sigma_z \rangle = - \tanh (\omega_q/2 T) < 0$. This is justified at the level of the differential equation~\eqref{eq:deq_matrix_form} by the fact that the time-dependence of $\langle \sigma_z \rangle$ will decay due to relaxation processes in the long-time limit even in the presence of noise. A proof is given in Appendix~\ref{appx:perturbation_theory}.

While the model is general and describes any two-level system with fluctuating energy separation, we discuss the potential tunability of the noise couplings by considering the example of a charge qubit, i.e., a qubit that encodes the logical states into the bonding and anti-bonding orbital states of a charge in a double quantum dot. The qubit splitting is given by $\omega_q = \sqrt{\epsilon^2 +4 t_c^2}$, where $\epsilon$ is the detuning between the two dots and $t_c$ is the tunnel matrix element~\cite{Zhang_review2019, Chatterjee2021}. In the presence of charge noise due to imperfect gate voltages, the detuning fluctuates in time, $\epsilon \rightarrow \epsilon + \delta \epsilon(t)$, and we find $\lambda = \partial_{\epsilon} \omega_q \vert_{\delta \epsilon = 0} = \epsilon/ \omega_q$. On the other hand, the qubit-photon coupling strength has the parameter dependence $g \propto t_c / \sqrt{\epsilon^2 +4 t_c^2}$ and hence may also be affected by noise. To leading order the noise couples to $g$ with strength given by $\lambda' = \partial_{\epsilon} g \vert_{\delta \epsilon = 0} = - g\epsilon/ \omega_q^2$, and the ratio $\lambda'/\lambda  = - g/\omega_q \propto 1/\epsilon^2$ for $\epsilon \gg t_c$ can be controlled via the detuning. Therefore, if the system is operated at sufficiently large detunings, the assumption that only the energy separation is affected by noise is valid. Additionally, the prefactor of higher order noise terms $\delta X^{n \geqslant 2}$ in the expansion of the qubit energy fluctuations $\delta \omega_q(t) = \lambda \delta X + \sum_{n=2}^{\infty} \lambda^{(n)} \delta X^n/n!$ is suppressed in this regime. Specifically, one has $\lambda^{(n)} = \partial_{\epsilon}^n \omega_q  \vert_{\delta \epsilon = 0} \sim 1/\epsilon^{n+1}$, while the first order term $\lambda$ is asymptotically constant at large detunings.

Finally, we turn to the general framework for the description of the resonator transmission. The input field $\langle b_{\text{in} } \rangle$ appears in the Langevin equations~\eqref{eq:deq_matrix_form}, and standard input-output theory states the simple relation for the output field $\langle b_{\text{out} } (t)\rangle = - \sqrt{\kappa_2} \langle a(t)\rangle$~\cite{Collett1984,Gardiner1985,Burkard2020}. Consequently, the time-dependent transmission amplitude $A$ through the cavity is found to be
	\begin{align}
	\label{eq:transmission_input_output}
		A(t) = \frac{\langle b_{\text{out} }(t) \rangle}{\langle b_{\text{in} } \rangle} = - \sqrt{\kappa_2} \frac{\langle a(t) \rangle}{\langle b_{\text{in} } \rangle}.
	\end{align}
In order to compute the transmission amplitude $A$, it is therefore necessary to solve the system of coupled differential equations~\eqref{eq:deq_matrix_form} for the expectation value of the photon annihilation operator $\langle a(t) \rangle$. Since the dimension of the matrix $L$ is larger than one, a general solution for arbitrary time-dependent noise can only be given as a time-ordered series. To obtain tractable expressions that can lead to physical insights and allow us to extract noise characteristics, we apply time-dependent perturbation theory. In Secs.~\ref{sec:diagonalizable_L} and~\ref{sec:non-diagonalizable_L}, the perturbation parameter is the ratio of the maximum value of the noise taken until measurement and the smaller of the two decay rates $\kappa$ and $\gamma$. Hence, this approach is expected to be applicable in open cavities or in cavities containing quickly relaxing qubits. In Sec.~\ref{sec:perturbation_theory_in_g}, on the other hand, we expand the exact solution in the parameter $\vert g \vert /\text{max} \lbrace \vert \delta_0 \vert, \vert \kappa - \gamma \vert  \rbrace$, where $\delta_0 = \omega_c - \omega_q$ is the noise-free qubit-cavity detuning. The results obtained in this fashion are expected to be valid in the dispersive regime $\vert \delta_0 \vert \gg \vert g \vert$ or when there is a large discrepancy in the qubit and cavity decay rates.

\section{Perturbation theory in the noise}
\label{sec:perturbation_theory_in_noise}
The system of differential equations in~\eqref{eq:deq_matrix_form} may not be solved exactly for generic noise. In this section we obtain an approximate analytical solution by treating the noise as a perturbation, deriving the conditions of validity of the approach as we go along. Since the zeroth-order transmission amplitude may then be obtained by exponentiating the matrix $L$, we distinguish between the cases where $L$ is diagonalizable (Sec.~\ref{sec:diagonalizable_L}) and where it is not (Sec.~\ref{sec:non-diagonalizable_L}). We discuss and compare our findings in Sec.~\ref{subsec:discussion} and finally propose a scheme for extracting noise characteristics from the measured average transmission probability in Sec.~\ref{sec:extracting_noise_characteristics}.

\subsection{Diagonalizable system matrix}
\label{sec:diagonalizable_L}
For $\delta_0 \neq 0$ or $\delta_0 = 0$ while $(\kappa-\gamma)^2 \neq -16 \langle \sigma_z \rangle g^2$, the non-Hermitian matrix $L$ is diagonalizable with eigenvalues
	\begin{align}
	\begin{split}
	&	l_{\pm} = \Gamma +  i \Delta \pm \Lambda, \; \; \; \Lambda = \frac{1}{4}\sqrt{\left( \kappa- \gamma + 2i \delta_0 \right)^2 + 16 \langle \sigma_z \rangle g^2 },
	\end{split}
	\end{align}
where $\Gamma = (\kappa + \gamma)/4$ and $\Delta = (\Delta_c + \Delta_q)/2$. In the absence of noise the differential equations~\eqref{eq:deq_matrix_form} can be decoupled by expressing the matrix $L$ in its eigenbasis. Treating the matrix $\delta L$ and therefore $\lambda \delta X(t)$ as a perturbation then allows us to solve for $\langle a(t) \rangle$ order by order in the noise (Appendix~\ref{appx:perturbation_theory}). Going up to quadratic order and using the limiting cases discussed in Appendix~\ref{appx:long_time_limit}, we find for the transmission amplitude as given by~\eqref{eq:transmission_input_output} at long times $\text{Re} (l_{\pm}) t \gg 1$,
	\begin{align}
	\label{eq:transmission_amplitude}
	\begin{split}
		 &\frac{A(t)}{A_0} = 1 +  \frac{\langle \sigma_z \rangle g^2}{\Lambda \left( i \Delta_q + \gamma/2 \right)} \bigg[  \frac{i}{2}   \left( f_+(t) - f_- (t)\right) \\
		 &\qquad \qquad +   \sum_{\pm} \frac{\gamma - \kappa - 2i \delta_0 \pm 4 \Lambda}{16 \Lambda} \left( f_{\pm \pm}(t)  - f_{\mp \pm} (t) \right)  \bigg],
	\end{split}
	\end{align}
where	
	\begin{align}
	\begin{split}
	\label{eq:A0_fp_fpq}
		A_0 & = - \frac{\sqrt{\kappa_1 \kappa_2}}{i \Delta_c + \kappa/2 - \langle \sigma_z \rangle g^2/(i \Delta_q + \gamma/2)}, \\
		f_p(t) & = \lambda e^{ - l_p t } \int_0^t  d t_1 \delta X( t_1)  e^{ l_p t_1} , \\
		f_{p q}(t) & =   \lambda e^{ - l_p t } \int_0^t  d t_1 \delta X( t_1 )  e^{l_p   t_1} f_q(t_1).
	\end{split}
	\end{align}
Here, $A_0$ is the zeroth-order transmission which would be the steady state in an noise-free system, and the functions $f_p$ and $f_{pq}$ with $p,q \in \lbrace \pm \rbrace$ are the first- and second-order noise corrections, respectively. They depend on time, and hence a true steady state is not reached for a single noise realization. While it may appear that the limit $\Lambda \rightarrow 0$ could lead to divergences, a closer analysis reveals that the expressions $f_+ - f_-$ and $ \sum_{\pm} \pm ( f_{\pm \pm}  - f_{\mp \pm}) $ only have terms $\Lambda^n$ with $n \geq 1$, i.e., no constant term, and the expression $\sum_{\pm} ( f_{\pm \pm}  - f_{\mp \pm} )$ only has terms $\Lambda^n$ with $n \geq 2$. These are precisely the powers needed to cancel the negative powers of $\Lambda$ in~\eqref{eq:transmission_amplitude}, and hence the limit $\Lambda \rightarrow 0$ exists and is well behaved (Appendix~\ref{appx:limit_lambda_zero}). Note, however, that at $\Lambda = 0$, the matrix $L$ is non-diagonalizable, and must be treated in a different approach (Sec.~\ref{sec:non-diagonalizable_L}).

As can be seen from Eq.~\eqref{eq:transmission_amplitude} the noise integrals serve as the perturbation parameters, and at long times, $\text{Re} (l_{\pm}) t \gg 1$, they are bounded by
	\begin{align}
	\label{eq:noise_intergal_bounds}
	\begin{split}
		\vert f_p \vert &   \leqslant \vert f^{\text{max}}_p \vert = \frac{ \lambda \vert \delta X_{\text{max}} \vert}{\Gamma + p \text{Re} (\Lambda)\vert_{\delta_0=0}}, \\
		\vert f_{pq} \vert & \leqslant  \vert f^{\text{max}}_{pq} \vert = \vert f^{\text{max}}_p \vert \vert f^{\text{max}}_q \vert,
	\end{split}
	\end{align}
where $\vert \delta X_{\text{max}} \vert = \vert \text{max} \lbrace \delta X(\tau) : \tau \in [0,t] \rbrace \vert$ is the maximal value taken by the noise until measurement at time $t$. The regime in which the perturbation expansion is valid is therefore characterized by $ \vert f^{\text{max}}_p \vert  \ll 1$ for $p \in \lbrace \pm \rbrace$. Note that this condition is independent of the probe frequency $\omega_p$ and the qubit-cavity detuning $\delta_0$ which are the control parameters of choice in transmission experiments. Physically, the condition $ \vert f^{\text{max}}_p \vert  \ll 1$ can be understood as follows: in time-dependent perturbation theory the quality of the expansion is a function of time as the dynamical noise will move the solution away from a steady state. At long times the accuracy is determined solely by the interplay of noise and decay rates and independent of any oscillatory processes. The decay rates must then be large enough to tame the deviations from the (by then time-independent) zeroth-order solution induced by the noise.

Relevant quantities in input-output measurements include the transmission $\vert A \vert$ and the transmission probability $\vert A \vert^2$. In the literature both $\vert A \vert$ and  $\vert A \vert^2$ are used, and there is no universal convention as to which of the two quantities is used to characterize cavity transmission. Clearly, this is justified by the very simple relation between the two quantities in noise-free systems. However, when a noise average $\llangle \cdot \rrangle$ is involved, one must be cautious and include the variance of $\vert A \vert$, $\llangle \vert A \vert^2 \rrangle = \llangle \vert A \vert \rrangle^2 + \text{ Var} (\vert A \vert)$. In Ref.~\cite{Mutter2022} and Sec.~\ref{sec:perturbation_theory_in_g} of this paper the variance vanishes to the order of interest in perturbation theory, and the transmission and transmission probability can be used interchangeably. In contrast, the variance does not vanish in general when performing a perturbation expansion in the noise parameter as we do in this section. Hence, we must decide on using one figure of merit, and we choose the averaged transmission probability $\llangle \vert A \vert^2 \rrangle$. This is because we aim to extract noise features from measurable convolutions of the noise power spectral density with an integration kernel, and we find that the mean $\llangle \vert A \vert \rrangle$ and variance $\text{ Var} (\vert A \vert)$ of the transmission both contain a term that cannot be written as a sum over such convolutions (Appendix~\ref{appx:mean_and_variance}), while this problematic term is not present in the averaged transmission probability.

Squaring the transmission and averaging over many noise configurations is a cumbersome task, and it contains the additional complication that the radicand in $\Lambda$ has both real and imaginary parts. If the random process governing the noise dynamics is assumed to be stationary, the mean of the noise is time independent and any unconditional joint probability distribution is invariant under time translations. In this case the general solution has the form
	\begin{align}
	\label{eq:transmission_convolutions}
			& \frac{\llangle \vert A \vert^2 \rrangle}{\vert A_0 \vert^2} = 1 + \frac{\langle \sigma_z \rangle g^2}{4 \Delta_q^2 + \gamma^2} \left[ \lambda \varrho \llangle  \delta X \rrangle + \lambda^2  \sum_{j=1}^5 \psi_j \mathcal{C}_j(\Delta) \right],
	\end{align}
where $\vert A_0 \vert$ is the absolute value of the well-known steady-state transmission amplitude~\eqref{eq:A0_fp_fpq} in the absence of noise. The corrections to the noise-free steady-state solution are twofold: The term linear in the qubit-noise coupling strength $\lambda$ is only present for biased noise, i.e., fluctuations with non-zero mean $\llangle \delta X \rrangle$. The effect of a non-zero noise mean can be thought of as a shift in the qubit energy splitting, $\omega_q \rightarrow \omega_q + \lambda \llangle \delta X \rrangle$, as can be shown by expanding $\vert A_0 \vert^2$ to leading order in $\lambda \llangle \delta X \rrangle$, and hence we set $\llangle \delta X \rrangle = 0$ in the remainder of this paper. On the other hand, the term quadratic in $\lambda$ is present even for zero-mean noise and describes the effect of the noise correlations. The functions $\varrho$ and $ \psi_j$ depend on the system parameters in a rather complicated fashion and are displayed in Appendix~\ref{appx:psi_functions}. Finally, the quantities $\mathcal{C}_j (\Delta)$ are convolutions,
	\begin{align}
	\label{eq:define_convolution}
		\mathcal{C}_j (\Delta) = \left( S \star K_j \right)(\Delta)= \frac{1}{2 \pi} \int_{-\infty}^{\infty} S(\omega) K_j(\Delta - \omega) d \omega.
	\end{align}
Here, $S(\omega)$ is the noise power spectral density defined as the Fourier transform of the noise autocorrelator, and the integration kernels read
	\begin{align}
	\label{eq:convolution_kernels}
	\begin{split}
		K_1(\Delta) & = \prod_{\pm} \frac{1}{[\Gamma \pm \text{Re} (\Lambda)]^2 + [\Delta \pm \text{Im} (\Lambda)]^2}, \\
		K_{2/3}(\Delta) & = \frac{1}{[\Gamma \pm \text{Re} (\Lambda)]^2 + [\Delta \pm \text{Im} (\Lambda)]^2}, \\
		K_{4/5}(\Delta) & = \frac{\Delta \pm \text{Im} (\Lambda)}{[\Gamma \pm \text{Re} (\Lambda)]^2 + [\Delta \pm \text{Im} (\Lambda)]^2},
	\end{split}
	\end{align}
where the positive (negative) sign belongs to the even (odd) index in the latter two equations. A notable special case is $\delta_0 = 0$, $(\kappa- \gamma)^2 > - 16 g^2 \langle \sigma_z \rangle$ for which $\text{Im} (\Lambda) =0$. The kernel $K_1$ can then be written as a linear combination of the kernels $K_2$ and $K_3$ by means of a partial fraction decomposition. Since the coefficients are independent of $\Delta$, we find for the sum of convolutions,
		\begin{align}
	\label{eq:transmission_convolutions_reduced}
			&  \sum_{j=1}^5 \psi_j \mathcal{C}_j(\Delta) = \sum_{j=2}^5 \tilde{\psi}_j \mathcal{C}_j(\Delta),
	\end{align}
where the functions $\tilde{\psi}_j$ relate to the original functions $\psi_j$ via
	\begin{align}
	\label{eq:psi_tilde_functions}
		\tilde{\psi}_{2/3} = \psi_{2/3} \mp \frac{\langle \sigma_z \rangle g^2}{ \text{Re}(\Lambda) \Gamma}, \quad \tilde{\psi}_{4/5} = \psi_{4/5}.
	\end{align}
Reducing the number of convolutions in this way can ease their experimental extraction as we detail in Sec.~\ref{sec:extracting_noise_characteristics}.

\subsection{Non-diagonalizable system matrix}
\label{sec:non-diagonalizable_L}

For $\delta_0 = 0$ and $( \kappa - \gamma)^2 = - 16 g^2 \langle \sigma_z \rangle$ one has $\Lambda = 0$, and the eigenvalues of $L$ become degenerate, $l_+ = l_- \equiv l$. Since in this case the algebraic multiplicity exceeds the geometric multiplicity, the matrix is no longer diagonalizable and may only be brought into Jordan normal form,
	\begin{align}
	\label{eq:matrix_Jordan_normal_form}
		L_J  = \begin{pmatrix}
			l/g  & 1 \\
			0 & l/g 
		\end{pmatrix},
		\quad l = \Gamma + i \Delta,
	\end{align}
where now $\Delta = \Delta_c = \Delta_q$. In the generalized eigenbasis, the system of differential equations is then only partially decoupled with one equation still mixing the dependent variables. Nevertheless, the Langevin equations~\eqref{eq:deq_matrix_form} can be solved within perturbation theory, and the transmission amplitude up to quadratic order in the noise $\delta X(t)$ reads
	\begin{align}
	\label{eq:transmission_critical_T}
		& \frac{A(t) }{A_0} = 1 - \frac{ \langle \sigma_z \rangle g^2}{ i \Delta + \gamma/2} \bigg[ i I_1(t) +  I_2(t) + \frac{\kappa - \gamma}{4} I_3(t)\bigg],
		\end{align}
with the noise integrals		
		\begin{align}
		\label{eq:noise_integrals_critical}
		 I_1(t) & = \lambda e^{-l t} \int_0^t dt_1 \int_0^{t_1} dt_2 \delta X(t_2) e^{l t_2}, \nonumber \\
		 I_2 (t) & = \lambda^2 e^{-l t} \int_0^{t} dt_1 \int_0^{t_1} dt_2 \delta X(t_2) \int_0^{t_2} dt_3 \delta X(t_3) e^{l t_3}, \nonumber \\
		 I_3(t) & =  \lambda e^{-l t} \int_0^{t} dt_1 \int_0^{t_1} dt_2 \delta X(t_2) e^{l t_2} I_1(t_2).
	\end{align}
The zeroth-order transmission $A_0$ is simply the standard steady-state solution~\eqref{eq:A0_fp_fpq} evaluated at the critical parameter settings. As before the noise integrals are the perturbation parameters, and in the long-time limit $\Gamma t \gg 1$ they are bound by 
	\begin{align}
	\label{eq:noise_integrals_bound_critical_case}
	\begin{split}
		 \Gamma \vert I_1 \vert & \leqslant I_{\text{max}} = \frac{ \lambda \vert X_{\text{max}} \vert}{\Gamma}, \\
		 \Gamma \vert I_2 \vert & \leqslant I_{\text{max}}^2, \quad  \Gamma^2   \vert I_3 \vert \leqslant I_{\text{max}}^2,
	\end{split}
	\end{align}			
and hence we require $\Gamma \gg \vert \delta X_{\text{max}} \vert$, motivated by the same physical considerations as in Sec.~\ref{sec:diagonalizable_L}. Again assuming the noise process to be stationary, the averaged long-time transmission probability reads
	\begin{align}
	\label{eq:ASTP_critical}
		 \frac{\llangle \vert A \vert^2 \rrangle}{\vert A_0 \vert^2} = 1 + \frac{ \langle \sigma_z \rangle g^2}{4 \Delta^2 +\gamma^2 } \left[ \lambda \varrho \llangle \delta X \rrangle + \lambda^2 \sum_{j=1}^3 \psi_j \mathcal{C}_j(\Delta) \right],
	\end{align}
where we include a non-zero mean only for completeness, and where in contrast to Sec.~\ref{sec:diagonalizable_L} the functions $\varrho$ and $\psi_j$ have compact forms, reading  $\varrho = -\psi_3$, $\psi_1 = 4 \langle \sigma_z \rangle g^2$ and
	\begin{align}
	\begin{split}
	\label{eq:psi_functions_critical}
		\psi_2 = \frac{ 8 \Gamma \Delta^2 - 2 (\Gamma^2 - \Delta^2)  \gamma}{(\Gamma^2 + \Delta^2)^2}, \; \psi_3 =  \frac{8 \Delta (\Gamma \gamma + \Gamma^2 - \Delta^2)}{(\Gamma^2 + \Delta^2)^2}.
	\end{split}
	\end{align}
Finally, the convolutions $\mathcal{C}_j$ are as defined in Eq.~\eqref{eq:define_convolution}, and the integration kernels read
	\begin{align}
	\label{eq:convolution_kernels_critical}
	\begin{split}
		 K_1(\Delta) & = \frac{1}{\left( \Gamma^2 + \Delta^2   \right)^2}, \quad K_2(\Delta) = \frac{\kappa \Gamma^2 + \gamma \Delta^2 }{ \left( \Gamma^2 + \Delta^2 \right)^2   }, \\  
		 K_3(\Delta) & = \Delta \frac{\Gamma^2 + \Delta^2 + (\kappa^2 - \gamma^2)/8}{\left( \Gamma^2 + \Delta^2   \right)^2}.
	\end{split}
	\end{align}
We now turn to discuss and compare our findings of Secs.~\ref{sec:diagonalizable_L} and~\ref{sec:non-diagonalizable_L}.

	\begin{figure*}
		\includegraphics[scale=0.31]{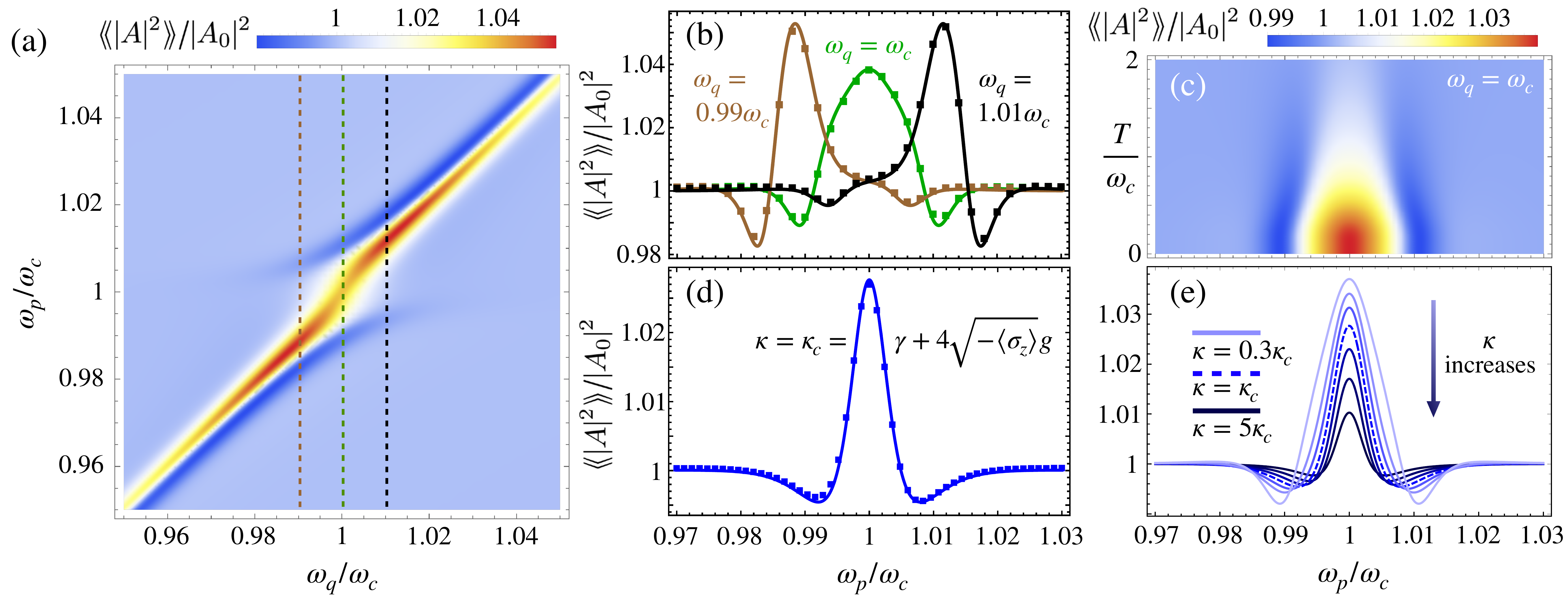}
		\caption{The normalized averaged long-time transmission probability. Noise corrections are visible whenever $\llangle \vert A \vert^2 \rrangle/\vert A_0 \vert^2 \neq 1$. (a) $\llangle \vert A \vert^2 \rrangle /\vert A_0 \vert^2$ as a function of the qubit frequency $\omega_q$ and the probe frequency $\omega_p$. The effect of the noise is most pronounced at the qubit-probe resonance $ \omega_q = \omega_p$. (b) Line cuts at different detunings as indicated by the dashed lines in (a). Analytical results (solid lines) are drawn according to Eq.~\eqref{eq:transmission_convolutions}, while the numerical results are obtained by averaging over $10^3$ solutions of the Lindblad master equation that corresponds to the exact Langevin equations. We include normally distributed quasistatic noise with zero mean and standard deviation $\sigma = \delta X_{\text{rms}}$, and allow for up to $12$ photons in the cavity. (c) Temperature dependence of the averaged transmission probability. At high temperatures the noise corrections are washed out. (d) $\llangle \vert A \vert^2 \rrangle /\vert A_0 \vert^2$ at the critical parameter settings, $\Lambda = 0$, i.e., $\kappa_c  = \gamma \pm 4 g \sqrt{- \langle \sigma_z \rangle}$ for $\kappa \gtrless \gamma$. Analytical results (solid lines) are drawn according to Eq.~\eqref{eq:ASTP_critical}, while the numerical results (squares) are obtained as in (b). (e) Comparison between the cases of a diagonalizable and non-diagonalizable system matrix $L$. We find that the transmissions for $\kappa \neq \kappa_c $ smoothly transition to the transmission at $\kappa = \kappa_c$ (dashed line). We set $\gamma t = 10^3$, $\lambda \delta X_{\text{rms}} =10^{-3} \omega_c $, $g = 0.01 \omega_c$, $\gamma_1 = 0.01 \omega_c$, $\gamma_{\varphi} = 0.005 \omega_c$ and $\kappa_{\text{int}} = 0$ in all panels, $\omega_q/T = 10$ in (a), (b), (d), (e), $\kappa = 0.01 \omega_c$ in (a)-(c) and $\delta_0 = 0$ in (c)-(e).}
		\label{fig:quasistatic_noise}
	\end{figure*}

\subsection{Discussion}
\label{subsec:discussion}
The expressions in Eqs.~\eqref{eq:transmission_convolutions} and \eqref{eq:ASTP_critical} are two of the main analytical results of this paper. They describe the noise corrections to the long-time transmission probability for diagonalizable and non-diagonalizable system matrices $L$, respectively. A few remarks applying to both results are in order. (i) While the transmission probability for a single noise realization depends on time, this time dependence vanishes upon averaging over many measurements. Thus, the averaged transmission probability resembles a steady state. (ii) In contrast to the transient transmission discussed in Ref.~\cite{Mutter2022}, the signature of the noise is washed out at high temperatures, $\langle \sigma_z \rangle_{t\rightarrow\infty} \rightarrow 0$, i.e., when the thermal energy strongly exceeds the qubit splitting. (iii) On the other hand, in contrast to the transient signal, preparation of a coherent superposition with $\langle \sigma_- \rangle_{t=0} \neq 0$ as an initial state is not required here.
(iv) The assumption that the random process is stationary and the condition $\delta X_{\text{max}} \ll \Gamma$ are the only restrictions on the noise in this section. The results obtained for the transmission are general in the sense that we do not have to make any assumptions on the form of the noise apart from the requirement that the Fourier transform of the correlator, i.e., the spectral density must exist. In particular, we do not have to make the Gaussian approximation. This can be advantageous since the assumption that the random variable is normally distributed can be violated for noise types relevant to quantum information processing systems, as in the case of $1/f$ noise~\cite{Kogan1996} or for qubits perturbed by a single charge fluctuator~\cite{Shalak2022arXiv}. One should note, however, that the underlying random process may not be characterized completely by the spectral density $S$ if the Gaussian approximation is not valid, and higher-order polyspectra must be taken into account. (v) The functional form of the second-order correction terms is a sum of convolutions of the spectral density with an integration kernel. By measuring any of these convolutions, one may use the convolution theorem to extract the power spectral density $S$. In Sec.~\ref{sec:extracting_noise_characteristics} we elaborate on this scheme.

To illustrate the procedure and the validity of our expressions, we consider quasistatic noise, i.e., noise which is constant within one measurement but fluctuates between measurements. The noise autocorrelator is constant, and as a result one has $S(\omega ) = 2 \pi \delta X_{\text{rms}}^2 \delta (\omega)$, where $\delta X_{\text{rms}} = \sqrt{\llangle \delta X^2 \rrangle}$ is the root mean square of the noise and $\delta (\omega)$ is the Dirac delta distribution. We remark that in this case one may alternatively solve the Langevin equation for a single measurement exactly, and in the long-time limit one finds $A_0(\Delta_q) \rightarrow A_0(\Delta_q + \lambda \delta X)$ with $A_0$ as given in~\eqref{eq:A0_fp_fpq}. To average the transmission probability, one can expand the expression in orders of $\lambda \delta X/\Gamma$, and we find that both approaches agree. A comparison between the numerical results obtained by averaging over $10^3$ solutions to the exact Langevin equations allowing for up to $12$ photons in the resonator and the analytical results [Eqs.~\eqref{eq:transmission_convolutions} and~\eqref{eq:ASTP_critical}] is displayed in Fig.~\ref{fig:quasistatic_noise}. 

The figure shows the dependence of the normalized averaged long-time transmission probability on the experimentally relevant parameters $\omega_p$, $\omega_q$ and $T$. As can be seen from Fig.~\figref[(a)]{quasistatic_noise}, the fingerprints of the noise are most prominent along the resonance line $\Delta_q = 0$, i.e., when the probe frequency matches the qubit frequency, a direct consequence of the common prefactor of the correction terms in Eqs.~\eqref{eq:transmission_convolutions} and~\eqref{eq:ASTP_critical}. We generally find excellent agreement between our analytical expressions and numerical results [Fig.~\figref[(b)]{quasistatic_noise}]. In Fig.~\figref[(c)]{quasistatic_noise} we show the transmission as a function of temperature and the probe frequency at the qubit-cavity resonance $\delta_0 = 0$. At high temperatures, $\omega_q T \gtrsim 1$, the noise features are washed out, a common feature of all long-time noise corrections to the averaged transmission probability (see also Fig.~\ref{fig:transmission_diespersive_regime} in Sec.~\ref{sec:perturbation_theory_in_g}) and in contrast to the transient case~\cite{Mutter2022}. In Fig.~\figref[(d)]{quasistatic_noise} we compare the analytical and numerical results at the critical parameter settings, $\delta_0 = 0$ and $( \kappa - \gamma)^2 = - 16 g^2 \langle \sigma_z \rangle$, where the system matrix becomes non-diagonalizable, and we find excellent agreement also in this case. Finally, Fig.~\figref[(e)]{quasistatic_noise} shows the transmissions in a region around the critical parameter settings. The two distinct solutions from Secs.~\ref{sec:diagonalizable_L} and \ref{sec:non-diagonalizable_L} transition into each other smoothly, and this statement is proven in Appendix~\ref{appx:limit_lambda_zero}. Hence, the critical point in the parameter space is only of mathematical nature and does not possess any distinct physical properties in the long-time solution. Instead, the critical behaviour is expected to be present in the transient curve, the investigation of which is beyond the scope of this paper. Still, the results of Sec.~\ref{sec:non-diagonalizable_L} are valuable as they represent a simple special case in which the reported results can be cast in clear and relatively simple expressions.

\subsection{Extracting noise characteristics}
\label{sec:extracting_noise_characteristics}
In Secs.~\ref{sec:diagonalizable_L} and~\ref{sec:non-diagonalizable_L} we demonstrated that the averaged long-time transmission probability receives second-order noise corrections in the form of convolutions of the power spectral density with an integration kernel. The aim of this section is to investigate how the power spectral density can be extracted from these measurable convolutions, and how experiments should be designed in terms of the parameter settings to simplify the procedure.

Given the existence of the Fourier transforms of the functions $S$ and $K_j$, a convolution of the form $\mathcal{C}_j (\Delta) = (S \star K_j)(\Delta)$ as displayed in Eq.~\eqref{eq:define_convolution} may be formally inverted by using the convolution theorem. Let 
	\begin{align}
		\tilde{F}(\tau) = \frac{1}{2\pi} \int_{-\infty}^{\infty} F(\Delta) e^{-i \Delta \tau} d \Delta
	\end{align}
denote the Fourier transform of a given function $F(\Delta)$ with respect to the conjugate variable $\tau$. The convolution theorem then states $\tilde{\mathcal{C}}_j(\tau) = \tilde{S} (\tau) \tilde{K}_j (\tau)$, and hence one may recover the power spectral density at the frequency $\Delta$ by returning to the frequency domain with the Fourier back transform,
	\begin{align}
	\label{eq:S_via_convolution}
		S(\omega) = \int_{-\infty}^{\infty} \tilde{\mathcal{C}}_j(\tau) \tilde{K}_j^{-1} (\tau) e^{i \omega \tau} d \tau.
	\end{align}
The Fourier transforms $\tilde{K}_j (\tau)$ of the integration kernels appearing in Eq.~\eqref{eq:convolution_kernels} of Sec.~\ref{sec:diagonalizable_L} and Eq.~\eqref{eq:convolution_kernels_critical} of Sec.~\ref{sec:non-diagonalizable_L} exist and may be computed explicitly using the residue theorem as is shown in Appendix~\ref{appx:Fourier_transform_kernels}. As a result, the spectral density can be obtained from~\eqref{eq:S_via_convolution} once the convolutions $\mathcal{C}_j(\Delta)$ with $\Delta = \omega_c - \omega_p - \delta_0/2$ have been extracted from experiment. Since the probe frequency $\omega_p$ and the qubit-cavity detuning $\delta_0$ are tunable control parameters in transmission experiments, one may record $\mathcal{C}_j(\Delta)$ by sweeping one of these parameters. Note, however, that in order to compute the Fourier transform $ \tilde{\mathcal{C}}_j(\tau)$, the range of $\Delta$ must be sufficiently large to guarantee a pre-determined degree of accuracy. 

It is clear that, given Eq.~\eqref{eq:S_via_convolution} and our knowledge of the kernels $K_j$ and their Fourier transforms, the extraction of noise characteristics boils down to extracting the convolutions $\mathcal{C}_j$ as a function of $\Delta$. The expression for the averaged transmission probability is rather complicated, and the experimental extraction of the convolutions is a challenging task. Some special cases can simplify the expressions for the functions $\psi_j$ (see Appendix~\ref{appx:psi_functions}): If $\kappa = \gamma$ then the real part of $\Lambda$ vanishes, $\text{Re} (\Lambda)  = 0$. The same is true for $\delta_0 = 0$ and $(\kappa - \gamma)^2 < - 16 \langle \sigma_z \rangle g^2$. On the other hand, when $\delta_0 = 0$ and $(\kappa - \gamma)^2 > - 16 \langle \sigma_z \rangle g^2$, then the imaginary part of $\Lambda$ vanishes, $\text{Im} (\Lambda)  = 0$. In this case the number of convolutions in the noise corrections to the transmission can be reduced from five to four as was shown in Sec.~\ref{sec:diagonalizable_L}. Note that at $\delta_0 =0$ one can still vary $\Delta$ by changing the probe frequency $\omega_p$. In general, when either $\text{Re} (\Lambda) = 0$ or $\text{Im} (\Lambda) = 0$, the expressions simplify considerably, the simplest case being $\text{Re} (\Lambda) = \text{Im} (\Lambda) = 0$ as discussed in Sec.~\ref{sec:non-diagonalizable_L}, in which the number of convolutions is reduced from five to three. To further simplify matters, one may note that $\vert \psi_1 \mathcal{C}_1 \vert, \vert \psi_2 \mathcal{C}_2 \vert \ll \vert \psi_3 \mathcal{C}_3 \vert$ for all $\Delta$ except near the resonance at $\Delta = 0$. Outside this narrow region, $\psi_3 \mathcal{C}_3 (\Delta)$ can to a good approximation be extracted directly from the measured transmission.

When $\Lambda \neq 0$, it would be elegant to extract the convolutions $\mathcal{C}_j$ by recording for every value of $\Delta$ the sum $\mathcal{M}_i \equiv \mathcal{M} (\xi_i) = \sum_{j=1}^N \psi_j(\xi_i)  \mathcal{C}_j$ appearing in Eq.~\eqref{eq:transmission_convolutions} at $N \in \lbrace 4,5 \rbrace$ different parameter settings $\xi_i$ and solve the system of linear equations 
	\begin{align}
	\label{eq:system_of_linear_equations}
		\mathcal{M}_i = \sum_{j=1}^N \psi_{ij} \mathcal{C}_j, \; \psi_{ij} \equiv \psi_j(\xi_i) , \; i \in \lbrace 1, \dots , N \rbrace,
	\end{align}
for $\mathcal{C}_j$. There are two requirements on the set of distinct parameter configurations $\lbrace \xi_i \rbrace $ for this to be possible: (i) Firstly, it must be chosen such that the convolutions are unchanged, i.e., $ \mathcal{C}_j (\xi_k) = \mathcal{C}_j(\xi_l)$ for all $j,k,l \in \lbrace 1, \dots ,N \rbrace$. (ii) Secondly, the parameters must be chosen such that the matrix $\psi_{ij}$ is invertible, i.e., $\det \psi \neq 0 $. Unfortunately, it is impossible to satisfy both these conditions given the dependence of the functions $\psi_j$ and $\mathcal{C}_j$ on the system parameters. To see this, note that the convolutions~\eqref{eq:convolution_kernels} can only be kept constant as required in (i) by varying the parameters such that $\Lambda$ is constant. However, in this case the functions $\psi_j$ will be at most quadratic in the control parameter (e.g. in the detuning $\delta_0$). The equation $\sum_j \alpha_j \boldsymbol{\psi}_j = 0$ with $\boldsymbol{\psi}_j = (\psi_j(\xi_1), \dots, \psi_j (\xi_N))$ then only imposes three constraints on the $N > 3$ constants $\alpha_j$, resulting in an underdetermined homogeneous system of linear equations which always has a non-trivial solution according to the Rouch\'{e}-Capelli theorem. Hence, the vectors $\boldsymbol{\psi}_j$ are linearly dependent, and the determinant $\det \psi$ vanishes, violating condition (ii).

\begin{figure}
	\includegraphics[scale=0.3]{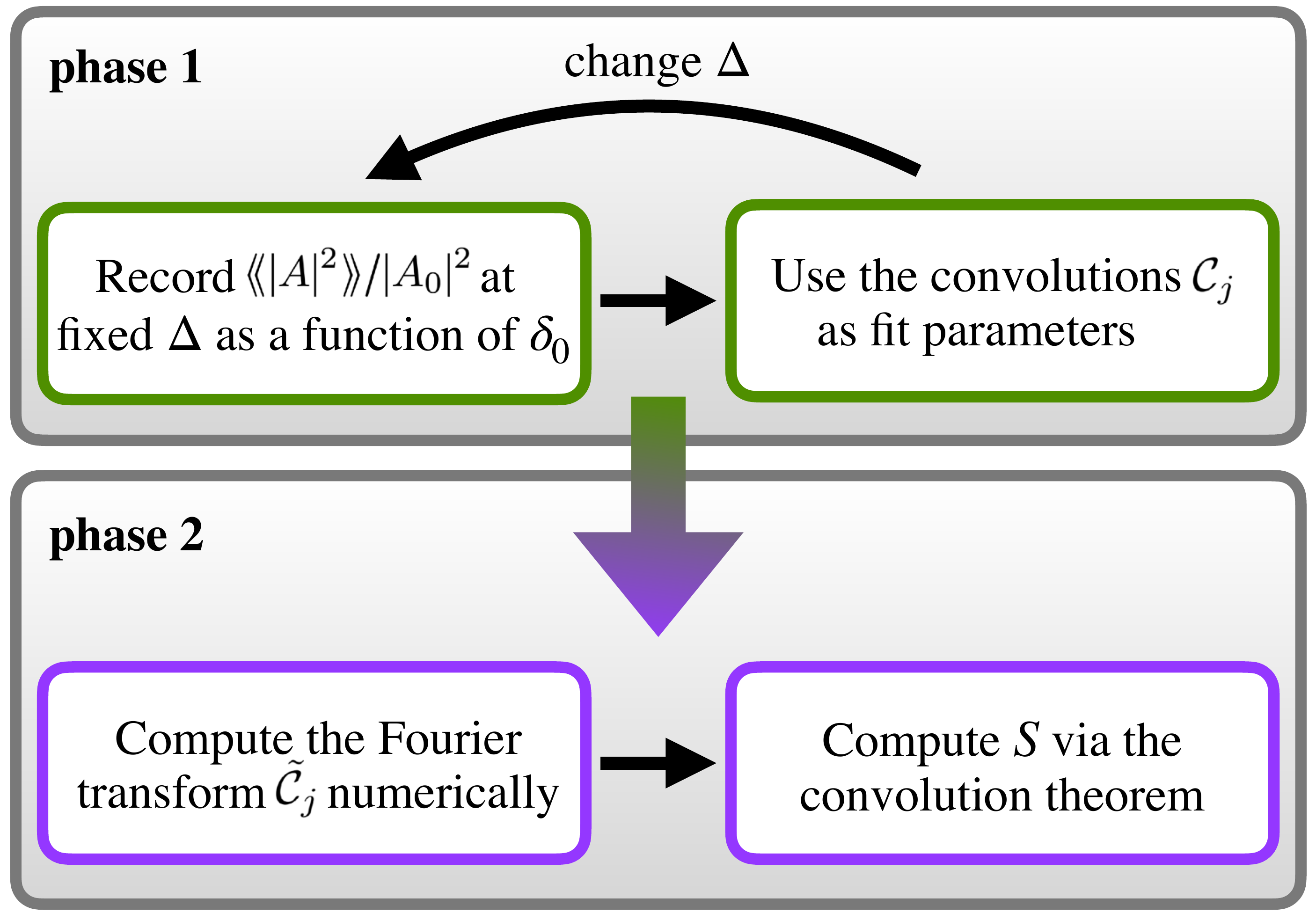}
		\caption{Schematic of the procedure to determine the power spectral density $S$. In the first phase, the averaged squared transmission is recorded for fixed $\Delta$ as a function of the detuning $\delta_0$ to extract the convolutions $\mathcal{C}_j$. After repeating this for many values of $\Delta$, the second phase is initiated. The Fourier transform of any of the convolutions is computed and finally the spectral density by applying the convolution theorem.}
		\label{fig:schematic_procedure}
	\end{figure}

As a consequence, one must resort to fitting the data to the expression~\eqref{eq:transmission_convolutions}. One can record the transmission at fixed $\Delta$ as a function of $\delta_0$, choosing the parameters such that the convolutions are constant when $\delta_0$ is varied. This corresponds to condition (i) above and is possible if in addition to the independent control over $\Delta$ and $\delta_0$ the qubit-photon coupling strength $g$ is tunable as well (Appendix~\ref{appx:extract_convolutions}). One may then fit the data to our expressions using the $\mathcal{C}_j$ as fit parameters and repeat this for many values of $\Delta$. A schematic of the procedure is shown in Fig.~\ref{fig:schematic_procedure}. 

While the approaches of extracting information on the noise affecting the qubit detailed in this section are laborious and challenging experimentally, this does not come as a surprise. The goal is to record an entire function, the power spectral density $S$, over the whole positive real axis or at least over a large frequency domain. Obtaining such a vast amount of information will always be demanding. Nevertheless, the schemes presented in this section are expected to be realizable in state-of-the-art experiments.

\section{Perturbation theory in the qubit-photon coupling}
\label{sec:perturbation_theory_in_g}
Finally, we work in the regime $\vert g \vert \ll  \text{max} \lbrace \vert \delta_0 \vert ,  \vert \kappa - \gamma \vert \rbrace $ in which one may treat the noise exactly. We now extend the results in Ref.~\cite{Mutter2022} by studying the long-time limit which is expected to be more easily accessible in experiment. After calculating the general long-time transmission in Sec.~\ref{subsec:long_time_transmission}, we discuss the analytically solvable case of white noise in Sec.~\ref{subsec:white_noise} and propose a way of extracting the power spectral density $S$ for arbitrary Gaussian noise in Sec.~\ref{subsec:extract_noise_characteristics_g}.
 
\subsection{Long-time transmission}
\label{subsec:long_time_transmission}
By treating $\varepsilon \equiv g/\text{max} \lbrace \vert \delta_0 \vert, \vert \kappa - \gamma \vert \rbrace$ as a small perturbation parameter, the system of differential equations~\eqref{eq:deq_matrix_form} can be decoupled and solved within perturbation theory. To leading order one has at long times $t \gg \text{max} \lbrace 1/\kappa, 1 / \gamma \rbrace$,
	\begin{align}
	\label{eq:long_time_transmission}
		\frac{A(t)}{A_{\infty}} = 1 + \langle \sigma_z \rangle g^2 \mathcal{N}(t), 
	\end{align}
where $A_{\infty}  = - \sqrt{\kappa_1 \kappa_2}/(i \Delta_c + \kappa/2)$ is the long-time transmission through an empty cavity, and where we introduce the noise integral
	\begin{align}
	\label{eq:second_order_noise_integral}
	\begin{split}
		\mathcal{N}(t) = e^{-i\Delta_c t - \kappa t/2} \int_0^t dt_2 e^{i (\Delta_c - \Delta_q) t_2 + (\kappa - \gamma)t_2/2}   \\
		\times \int_0^{t_2} dt_1 e^{i \Delta_q t_1 + \gamma t_1/2} e^{-i \lambda \mathcal{X}(t_1, t_2)},
	\end{split}
	\end{align}
containing the stochastic phase in the interval $[t_1,t_2]$,
	\begin{align}
	\label{eq:random_phase_interval}
		\mathcal{X}(t_1,t_2) = 	\mathcal{X}(t_2) - \mathcal{X}(t_1) = \int_{t_1}^{t_2}  \delta X(s) ds.
	\end{align}
It is shown in Appendix~\ref{appx:comparison} that the results for the transmission in Eq.~\eqref{eq:transmission_amplitude} of Sec.~\ref{sec:perturbation_theory_in_noise} and Eq.~\eqref{eq:long_time_transmission} of this section agree in the appropriate limit, i.e., to second order in both $ \delta X$ and $g$. In order to obtain a measurable quantity, we proceed to square the transmission amplitude~\eqref{eq:long_time_transmission} and average over many noise realizations. Since the term in the transmission linear in $g$ vanishes at long times, one has $\left \llangle \vert A(t) \vert^2 \right\rrangle = \left \llangle \vert A(t) \vert \right\rrangle^2$ up to and including quadratic order in $g$ (Appendix~\ref{appx:mean_and_variance}), and we find
	\begin{align}
	\label{eq:long_time_squared}
		\left \llangle \vert A(t) \vert \right\rrangle =  \vert A_{\infty} \vert \sqrt{1 + 2 \langle \sigma_z \rangle g^2 \text{Re}  \left\llangle \mathcal{N}(t) \right\rrangle},
	\end{align}
where $\left\llangle \mathcal{N}(t) \right\rrangle$ is the averaged noise integral (ANI),
	\begin{align}
	\label{eq:ANI_second_order}
	\begin{split}
		\left \llangle \mathcal{N}(t) \right \rrangle = e^{-i\Delta_c t - \kappa t/2} \int_0^t dt_2 e^{i (\Delta_c - \Delta_q) t_2 + (\kappa - \gamma)t_2/2}   \\
		\times \int_0^{t_2} dt_1 e^{i \Delta_q t_1 + \gamma t_1/2} \left \llangle e^{-i \lambda \mathcal{X}(t_1, t_2)} \right \rrangle.
	\end{split}
	\end{align}
One may most clearly see the noise signatures by considering the normalized deviation from the empty cavity transmission probability,
	\begin{align}
	\label{eq:normalized_deviation_long_time}
		\frac{\delta \llangle \vert A \vert^2 \rrangle}{\vert A_{\infty} \vert^2} = \frac{ \llangle \vert A \vert^2 \rrangle - \vert A_{\infty} \vert^2}{\vert A_{\infty} \vert^2} = 2 \langle \sigma_z \rangle g^2 \text{Re}  \left\llangle \mathcal{N}(t) \right\rrangle.
	\end{align}
In contrast to the transient transmission, the effect of the noise on the long-time transmission does not depend on the initial qubit state and is washed out at high temperatures, a feature we also observed in Sec.~\ref{sec:perturbation_theory_in_noise}. If the integration time exceeds the time on which the noise correlations decay, the central limit theorem predicts that the stochastic phase $\mathcal{X}(t_1,t_2) $ will be normally distributed at long times, and we find
	\begin{align}
	\label{eq:cumulant_expansion}
	\begin{split}
		&\left\llangle e^{-i \lambda \mathcal{X}(t_1,t_2)  }\right\rrangle = e^{- \lambda^2\left\llangle \mathcal{X}^2(t_1,t_2)  \right\rrangle /2 } \\
		& \qquad = \exp \left( - \frac{\lambda^2}{2 \pi} \int_{0}^{\infty} \frac{\sin^2 \left( \omega [t_2 - t_1]/2 \right)}{(\omega/2)^2}   S(\omega) d \omega \right).
	\end{split}
	\end{align}
Eq.~\eqref{eq:cumulant_expansion} is an obvious variant of the well-known expression describing the averaged qubit coherence $\llangle \langle \sigma_-(t) \rangle \rrangle$~\cite{Makhlin2004,Ithier2005,Chirolli2008, Bergli2009} with the integration time $t$ being replaced by the time difference $\Delta t = t_2 - t_1$.

\subsection{White noise}
\label{subsec:white_noise}
\begin{figure}
		\includegraphics[scale=0.2]{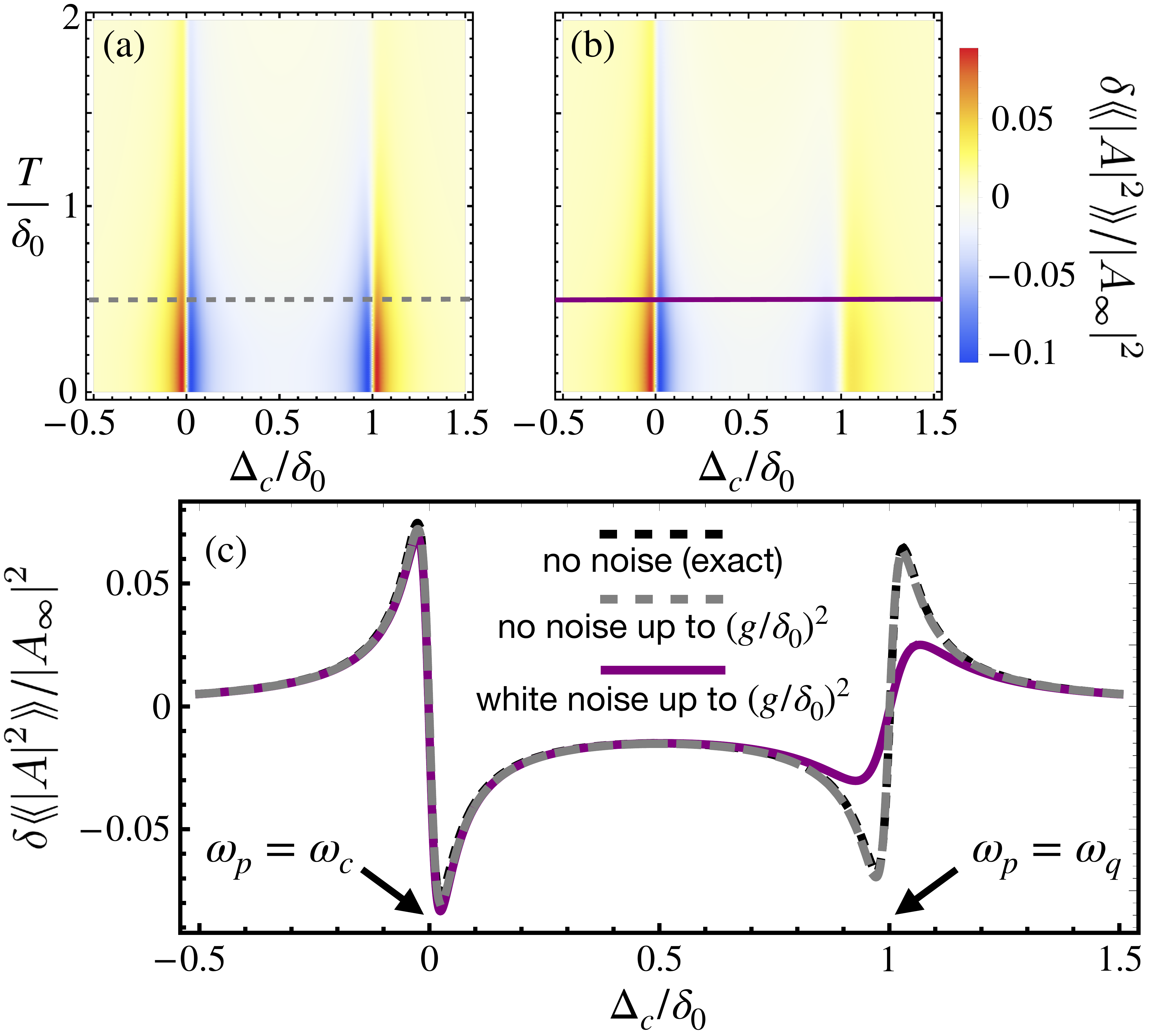}
		\caption{The normalized deviation from the empty cavity transmission probability $\delta \llangle \vert A \vert^2 \rrangle /\vert A_{\infty} \vert^2$. Panels (a) and (b) show $\delta \llangle \vert A \vert^2 \rrangle /\vert A_{\infty} \vert^2$ as a function of the cavity-probe detuning $\Delta_c$ and temperature $T$ in the noise-free case and in the presence of white noise of amplitude $S_0 = 0.1 \delta_0$, respectively. The noise affects the transmission features at the qubit frequency, $\Delta_c = \delta_0$, but not at the cavity frequency, $\Delta_c = 0$. At high temperatures, $T \gtrsim \omega_q$, the noise signatures are washed out. Panel (c) shows the line cuts at $T= 0.5 \delta_0$ as indicated by the horizontal lines  in (a) and (b). The parameters are set to $g = \kappa = 0.05 \delta_0$, $\kappa_{\text{int}} = 0$, $\gamma_1 = 0.05 \delta_0$, $\gamma_{\varphi} = 0.025 \delta_0$ and $\lambda = 0.9$.}
	\label{fig:transmission_diespersive_regime}
	\end{figure}

White noise is characterized by a constant spectral density, $S(\omega ) = S_0$ with noise amplitude $S_0$. In this case an exact solution for the real part of the ANI~\eqref{eq:ANI_second_order} appearing in Eq.~\eqref{eq:long_time_squared} may be obtained. For long times $t \gg \text{max} \lbrace 1/\kappa, 1/\gamma \rbrace$, one finds
	\begin{align}
	\label{eq:ANI_white_noise_real_part}
		\text{Re} \left\llangle  \mathcal{N}  \right\rrangle_w =  \frac{4 \kappa (\gamma +  \lambda^2 S_0)  - 16 \Delta_c \Delta_q}{\left[ 4\Delta_c^2 + \kappa^2 \right] \left[ 4\Delta_q^2 + (\gamma +  \lambda^2 S_0)^2 \right]}.
	\end{align}		
At $S_0 =0$,  Eq.~\eqref{eq:ANI_white_noise_real_part} reduces to the expression for the noise-free case and the squared transmission agrees with the expansion of the exact noise-free steady-state solution to second order in $g$. The Markovian white noise only renormalizes the decoherence rate from the Lindblad formalism which is also a Markovian theory, $\gamma \rightarrow \gamma +  \lambda^2 S_0$. As in Sec.~\ref{sec:perturbation_theory_in_noise} we note that while the transmission in a single-shot measurement will be a function of time, the averaged transmission is time independent, thus resembling a steady state. In Fig.~\ref{fig:transmission_diespersive_regime} we compare the normalized deviation from the empty cavity transmission probability in the presence of white noise to the noise-free case. As can be seen from the figure, the noise washes out the transmission features around the qubit-probe resonance $\omega_p = \omega_q$, while it does not affect the features at the cavity-probe resonance $\omega_p = \omega_c$, a consequence of the noise only affecting the two-level system. In the same figure we show that the noise traces disappear at high temperatures. This is a general observation in the long-time transmission and in stark contrast to the transient transmission which to leading order is temperature independent in signal strength.

\subsection{Extracting noise characteristics}	
\label{subsec:extract_noise_characteristics_g}
For generic noise types the ANI cannot be evaluated analytically. It is possible, however, to extract information about the power spectral density $S$ for arbitrary fluctuations if one works in the regime $ \vert g \vert \ll \vert \kappa - \gamma \vert$. Assume that the real part of the ANI has been recorded as a function of the noise coupling strength $\lambda$ (in a semiconductor charge qubit $\lambda$ can be controlled by the detuning) and the qubit-probe detuning $\Delta_q$ by transmission measurements from~\eqref{eq:normalized_deviation_long_time}. One may then consider the second derivative of the ANI in the long-time limit with $e^{- \gamma t}, e^{- \kappa t} \ll 1$,
	\begin{align}
	\label{eq:second_derivative}
	\begin{split}
			D_{\mathcal{N}}^2 &  := -\frac{d^2 \text{Re} \llangle \mathcal{N} \rrangle}{d \lambda^2} \bigg\vert_{\lambda = 0} =   \frac{2}{\pi} \text{Re}  \left( \int_{0}^{\infty} \frac{ S(\omega)}{ I(\omega)} d \omega \right), \\
			I (\omega) & =  \left( i \Delta_c +  \frac{\kappa}{2} \right) \left(i \Delta_q +  \frac{\gamma}{2} \right) \left[ \left(i \Delta_q + \frac{\gamma}{2} \right)^2 + \omega^2 \right].
	\end{split}
	\end{align}
Note that one may obtain the second derivative by recording $\text{Re} \llangle \mathcal{N} \rrangle$ for three values of $\lambda$ close to zero and fitting a parabola. Our aim in the following is to manipulate the expression~\eqref{eq:second_derivative} such that noise characteristics can be extracted from the quantity $D_{\mathcal{N}}^2$. By noting that
	\begin{align}
		\left(i \Delta_q + \frac{\gamma}{2} \right)^2 + \omega^2 = \prod_{\pm} \left[ i (\Delta_q \pm \omega) + \frac{\gamma}{2} \right],
	\end{align}
employing a partial fraction decomposition and using the symmetry of the spectral density for classical fluctuations in $\omega$, the expression~\eqref{eq:second_derivative} may be brought into the form
	\begin{align}
	\label{eq:second_derivative_2}
	\begin{split}
		D_{\mathcal{N}}^2(\Delta_q) =   \frac{\left[ \kappa \left( \gamma^2 - 4 \Delta_q^2 \right) - 8 \gamma \Delta_q \Delta_c \right] \text{Re} (\mathcal{C}(\Delta_q))}{(\kappa^2 + 4 \Delta_c^2)(\gamma^2 + 4 \Delta_q^2)^2} \\
		+ \frac{ \left[ 2 \Delta_c \left( \gamma^2 - 4 \Delta_q^2 \right) - 4 \kappa \Delta_q \gamma \right] \text{Im} (\mathcal{C} (\Delta_q)) }{(\kappa^2 + 4 \Delta_c^2)(\gamma^2 + 4 \Delta_q^2)^2}.
	\end{split}
	\end{align}
Here, $\mathcal{C} (\Delta_q ) = (S \star K)(\Delta_q)$ is a convolution as defined in Eq.~\eqref{eq:define_convolution} of the power spectral density with the kernel $K(\Delta_q) = 1/(i \Delta_q + \gamma/2)$. The real and imaginary parts of $\mathcal{C} (\Delta_q )$ can be extracted by recording $D_{\mathcal{N}}^2$ at fixed $\Delta_q$ but for two distinct values of $\Delta_c$ (this can be achieved by varying both $\omega_p$ and $\omega_q$) and solving the resulting system of linear equations (Appendix~\ref{appx:extract_convolutions_g}). The Fourier transform of the kernel is found analytically by applying the residue theorem,
	\begin{align}
	\label{eq:FT_kernel}
			\tilde{K}(\tau) = \frac{1}{2 \pi} \int_{-\infty}^{\infty} K(\Delta_q) e^{-i \Delta_q \tau} d \Delta_q = \begin{cases}  e^{ \gamma \tau/2} & \tau < 0 \\
			0 & \tau > 0
			\end{cases}.
	\end{align}	
$\tilde{K}(\tau) $ is discontinuous at $\tau = 0$ because $K(\Delta_q)$ is not absolutely integrable. However, it is square integrable, and so the Fourier transform exists. Finally, we may once again use the convolution theorem to solve for the power spectral density, 
	\begin{align}
	\label{eq:S_via_convolution_dispersive}
		S(\omega) = 2 \int_{-\infty}^0 \tilde{\mathcal{C}}(\tau)  \cos (\omega \tau)  e^{ -\gamma \tau/2} d \tau,
	\end{align}
where the inherited symmetry of $\tilde{S} (\tau)$ allows us to integrate only over negative $\tau$, avoiding a zero denominator when inserting the expression in Eq.~\eqref{eq:FT_kernel}. Eq.~\eqref{eq:S_via_convolution_dispersive} is the main result of this section. It states that it is possible to obtain the power spectral density for arbitrary noise from transmission measurements in the long-time limit by determining the convolution $\mathcal{C}(\Delta_q)$ via~\eqref{eq:second_derivative_2} and calculating its Fourier transform $\tilde{\mathcal{C}}(\tau)$. The range of the qubit-probe detuning $\Delta_q$ sets the accuracy with which $\tilde{\mathcal{C}}(\tau)$ may be computed.

It ought to be noted that the quantity~\eqref{eq:second_derivative} which forms the basis of our noise extraction scheme in this section requires knowledge of the ANI close to $\lambda = 0$, a point at which corrections due to a fluctuating qubit-photon coupling $g$ parametrized by the coupling strength $\lambda'$ may become important, for instance in a charge qubit at low detunings (see Sec.~\ref{sec:model}). In this case, the result in Eq.~\eqref{eq:S_via_convolution_dispersive} is only valid when the fluctuations in $g$ are absent or weak at $\lambda \sim 0$. On the other hand, when $\lambda$ is tuned via the tunnel coupling $t_c$, the condition $t_c \gg \epsilon$ needed to obtain values of $\lambda$ close to zero yields the scaling $\lambda'/\lambda \sim 1/t_c$ and hence guarantees $\lambda'/\lambda \ll 1$. In this case, spectral features can be inferred from the curvature of the real part of the ANI.

\section{Conclusions}
\label{sec:conclusions}

In summary, we develop a comprehensive theory describing the effect of dynamical noise affecting a general two-level system placed in a cavity on the long-time transmission. By solving the quantum Langevin equations in time-dependent perturbation theory, we demonstrate that the averaged long-time transmission probability receives noise corrections and that these corrections may be written as sums over convolutions containing the noise power spectral density $S$. We invert these relations using the convolution theorem, and obtain integral expressions for $S$ containing only measurable quantities. Within our model we are able to treat random fluctuations that lead to corrections in the long-time transmission of a few percent of the well-known steady-state transmission through a noise-free system. This effect is expected to be large enough to be resolvable in state-of-the-art experiments, and hence our results suggest the possibility of extracting noise characteristics from long-time transmission measurements.

Future research could explore the effect of gain in the cavity, e.g., pumping schemes realized by pulses applied to the system via the input field $\langle b_{\text{in}}(t)\rangle$. While such modifications complicate the experimental setup, they have the potential of revealing noise characteristics in cleaner functional forms. Additionally, to have a more complete description of the system and to enlarge the regime of validity of the model, one may take into account a fluctuating qubit-photon coupling constant $g$. Finally, it would be worthwhile to investigate the fingerprints of non-Gaussian noise on the cavity transmission and devise possible schemes to characterize the noise polyspectra.

\section{Acknowledgments}
We thank M. Benito for enlightening discussions on the form and validity of the quantum Langevin equations. This research is supported by the German Research Foundation [Deutsche Forschungsgemeinschaft (DFG)] under Project No.~450396347 and No.~425217212 - SFB 1432.

\appendix
\onecolumngrid

\section{Perturbation theory for differential equations and the time dependence of $\langle \sigma_z \rangle$}
\label{appx:perturbation_theory}

\subsection{General perturbation theory}

Consider a general system of inhomogeneous coupled differential equations in matrix form,
	\begin{align}
	\label{eq:DEQ_matrix}
		D_t \mathbf{x}(t) + \mathcal{L} \mathbf{x}(t) = \mathbf{b}(t),
	\end{align}
where $D_t$ is an unspecified operator containing derivatives with respect to the independent variable $t$, $ \mathcal{L}$ is a square matrix and $\mathbf{b}(t)$ an inhomogeneity. The goal is to solve for the vector of dependent variables $\mathbf{x}$ within perturbation theory. To this end, assume that we may separate the matrix $ \mathcal{L}$ into a part $L$ which we may diagonalize (or bring into Jordan normal form) and a part $\delta L$ which is of the order of a small perturbation parameter $\varepsilon \ll 1$, $ \mathcal{L} = L +  \delta L$. With the similarity transformation $\bar{L} = V L V^{-1}$, where $\bar{L}$ is $L$ in its (generalized) eigenbasis, we obtain for $V \neq V(t)$,
	\begin{align}
	\label{eq:DEQ_matrix_diagonal}
		D_t \bar{\mathbf{x}} (t) + \bar{L} \bar{\mathbf{x}} (t)  +  V \delta L V^{-1} \bar{\mathbf{x}} (t) = \overline{\mathbf{b}}(t),
	\end{align}
where $\bar{\mathbf{x}} = V \mathbf{x}$ and $\overline{\mathbf{b}} = V \mathbf{b}$. The solution to the transformed equation may be written as a perturbation series,
	\begin{align}
	\label{eq:solution_expansion}
		\bar{\mathbf{x}}(t) = \sum_{n=0}^{\infty}  \bar{\mathbf{x}}_n(t),
	\end{align}
where $\bar{\mathbf{x}}_n$ is of the order $n$ in the small parameter $\varepsilon$. Substituting the ansatz~\eqref{eq:solution_expansion} into Eq.~\eqref{eq:DEQ_matrix_diagonal} and collecting terms of the same order in $\varepsilon$, we obtain tractable equations for $\bar{\mathbf{x}}_n(t)$,
	\begin{align}
	\begin{split}
	\label{eq:DEQ_orders_of_epsilon}
		D_t \bar{\mathbf{x}}_0 (t) + \bar{L} \bar{\mathbf{x}}_0 (t) & = \overline{\mathbf{b}}(t), \\
		D_t \bar{\mathbf{x}}_n (t) + \bar{L} \bar{\mathbf{x}}_n (t) & = - V \delta L V^{-1} \bar{\mathbf{x}}_{n-1} (t)  , \quad n \geqslant 1.
	\end{split}
	\end{align}
These equations may now be solved to the desired order in $\varepsilon$. If the matrix $L$ is diagonalizable, the differential equations at each order decouple, while they are only partially decoupled if $L$ can only be brought into Jordan normal form. The solution to the original equation~\eqref{eq:DEQ_matrix} is given by
	\begin{align}
	\label{eq:solution_original_equation}
		\mathbf{x}(t) = V^{-1} \sum_{n=0}^N  \bar{\mathbf{x}}_n(t) + \mathcal{O} \left( \varepsilon^{N+1} \right),
	\end{align}
where $N$ is the order of truncation. In the main text we use this approach to solve the Langevin equations~\eqref{eq:deq_matrix_form} for the expectation value $\langle a(t) \rangle$, expanding in the parameters $\varepsilon = \vert f_{\pm} \vert$ (Eq.~\eqref{eq:noise_intergal_bounds}, Sec.~\ref{sec:perturbation_theory_in_g}) and $\varepsilon = \vert g \vert /\text{max} \lbrace \vert \delta_0 \vert, \vert \kappa - \gamma \vert \rbrace$ (Sec.~\ref{sec:perturbation_theory_in_g}) with $N=2$.

\subsection{Time dependence of $\langle \sigma_z \rangle$}

At this point one may consider the effect of a time dependence of $\langle \sigma_z \rangle$ in the Langevin equations given in Eq.~\eqref{eq:deq_matrix_form} of the main text. Writing $\langle \sigma_z \rangle = \langle \sigma_z^{\infty} \rangle + \xi(t) e^{- \bar{\gamma} t}$ with the steady-state solution $ \langle \sigma_z^{\infty} \rangle$ (which is assumed to be a thermal distribution in the main text), a characteristic decay rate $\bar{\gamma} >0$ and a function $\xi(t)$ bounded in $t$, we obtain the Langevin equations given in Eq.~\eqref{eq:deq_matrix_form} plus an additional time-dependent matrix term $ \sim \Xi(t) e^{- \bar{\gamma} t}$. Since $\xi(t)$ is bounded in $t$, so is the associated matrix $\Xi (t)$. In our perturbative approach, the time dependence of $\langle \sigma_z \rangle$ results in an additional term in the $n$th order equation for $\bar{\mathbf{x}} = V(\langle \sigma_- \rangle , \langle a \rangle)$ according to Eq.~\eqref{eq:DEQ_orders_of_epsilon} above,
	\begin{align}
	\begin{split}
		 \frac{d \bar{\mathbf{x}}_n (t)}{dt} + \bar{L} \bar{\mathbf{x}}_n (t) = - V \delta L V^{-1} \bar{\mathbf{x}}_{n-1} (t)  - e^{- \bar{\gamma} t} V  \Xi(t)  V^{-1} \bar{\mathbf{x}}_{n-1} (t),
	\end{split}
	\end{align}
where $\bar{L} = V L V^{-1}$ is the time-independent system matrix in its (generalized) eigenbasis. The general solution is
    \begin{align}
        \bar{\mathbf{x}}_n (t) = e^{-\bar{L} t} \left(\bar{\mathbf{x}}_0 (0) - \int_0^t e^{\bar{L} t'}  V \delta L (t') V^{-1} \bar{\mathbf{x}}_{n-1} (t')  dt' - \int_0^t e^{\bar{L} t'} e^{- \bar{\gamma} t'} V  \Xi(t')  V^{-1} \bar{\mathbf{x}}_{n-1} (t')  dt' \right).
    \end{align}
We will now show that the additional part due to the time-dependence of $\langle \sigma_z \rangle$,
    \begin{align}
        \boldsymbol{\eta }_n := - e^{-\bar{L} t} \int_0^t e^{\bar{L} t'}   e^{- \bar{\gamma} t'} V \Xi(t') V^{-1} \bar{\mathbf{x}}_{n-1} (t')  dt',
    \end{align}
vanishes in the long-time limit at all orders $n$ and hence exactly in the full long-time solution. Writing $\boldsymbol{\eta}_n = (\eta_n^1, \eta_n^2)$, $\bar{L} = \text{diag} (l_1,l_2)$ with $\text{Re}(l_{1,2}) > 0 $ and defining $\mathbf{z}_n(t) = V \Xi(t) V^{-1} \bar{\mathbf{x}}_{n-1} (t) $ with $\mathbf{z}_n = (z_n^1, z_n^2)$, one has for the components $i \in \lbrace 1,2 \rbrace$,
    \begin{align}
         \eta^i_n = - e^{- l_i t} \int_0^t z_n^i(t') e^{(l_i - \bar{\gamma})t'} dt'.
    \end{align}
The functions $\vert z_n^i(t) \vert$ are bounded in $t$ since all their constituents are, and we have
    \begin{align}
        0 \leqslant \vert  \eta^i_n \vert  \leqslant  \vert  z^i_{n,\text{max}} \vert e^{- \text{Re}(l_i)t } \int_0^t e^{(\text{Re}(l_i) - \bar{\gamma})t'} dt' = \begin{cases} 
        \vert z^i_{n,\text{max}} \vert \frac{e^{-\bar{\gamma} t} - e^{-\text{Re}(l_i) t}}{\text{Re}(l_i) - \bar{\gamma}} & \bar{\gamma} \neq \text{Re}(l_i) \\
         \vert z^i_{n,\text{max}}  \vert t e^{-\bar{\gamma} t}  &\bar{\gamma} = \text{Re}(l_i)
        \end{cases},
    \end{align}
where $\vert  z^i_{n,\text{max}} \vert = \text{max} \lbrace \vert  z_n^i(t') \vert : t' \in [0,t] \rbrace$. Since $\text{Re}(l_i), \bar{\gamma} >0$, the rightmost expression vanishes and $\eta^i_n \rightarrow 0$ in the long-time limit at all orders $n$. Hence, it is sufficient to consider the steady-state value of $\langle \sigma_z \rangle$ even at the level of the differential equations~\eqref{eq:deq_matrix_form}.

\section{The long-time limit}
\label{appx:long_time_limit}
To simplify the calculations when solving the quantum Langevin equations to second order in the noise, we note that since the absolute value squared $\vert \cdot \vert^2$ is a continuous function, one has
	\begin{align}
	\label{eq:limit_continous_function}
			\lim_{t \rightarrow \tau} \left\vert A(t) \right\vert^2 = \left\vert \lim_{t \rightarrow \tau} A(t) \right\vert^2
	\end{align}
if the limit $\lim_{t \rightarrow \tau} A(t)$ exists. It is therefore possible to take the long-time limit already in the expression for the transmission amplitude. When taking the limit, we use the estimate
	\begin{align}
	\label{eq:integral_limit}
		0 \leqslant I^{(1)}(\alpha, \beta) \equiv \left\vert e^{-\alpha t} \int_0^t  \delta X(t^{\prime}) e^{\beta t^{\prime}} d t^{\prime}  \right\vert \leqslant \left\vert e^{-\alpha t} \right\vert  \int_0^t  \left\vert \delta X(t^{\prime}) \right\vert  \left\vert e^{\beta t^{\prime}}\right\vert d t^{\prime} \leqslant \left\vert e^{-\alpha t} \right\vert  \int_0^t  \left\vert \delta X_{\text{max}} \right\vert  \left\vert e^{\beta t^{\prime}}\right\vert d t^{\prime},
	\end{align}
where $\delta X_{\text{max}} < \infty$ is the maximal value taken by the noise in the interval $[0,t]$ under consideration, and the real part of $\alpha$ is positive. When the real part of $\beta$ is zero, the rightmost expression is zero for $t \rightarrow \infty$ and hence the original integral $I^{(1)}$.  Writing $\alpha_R$ and $\beta_R$ for the real parts of $\alpha$ and $\beta$, respectively, we find for $\beta_R \neq 0$,
	\begin{align}
	\label{eq:integral_limit_2}
		0 \leqslant I^{(1)}(\alpha, \beta) \leqslant \left\vert \delta X_{\text{max}} \right\vert e^{-\alpha_R t} \int_0^t  e^{\beta_R t^{\prime}} d t^{\prime}   = \frac{\left\vert \delta X_{\text{max}} \right\vert }{\beta_R} \left( e^{(\beta_R - \alpha_R)t} - e^{-\alpha_R t} \right).
	\end{align}
At $t \rightarrow \infty$ the rightmost expression and hence the integral $I^{(1)}$ vanishes if $\beta_R < \alpha_R$.

A similar estimate holds for the second-order noise integral,
	\begin{align}
	\label{eq:integral_limit_second_order}
		0 \leqslant I^{(2)}(\alpha, \beta , \gamma) \equiv \left\vert e^{-\alpha t} \int_0^t d t^{\prime}  \delta X( t^{\prime} ) e^{\beta t^{\prime} } \int_0^{t^{\prime}} d t^{\prime \prime}  \delta X( t^{\prime \prime} ) e^{\gamma t^{\prime \prime} }  \right\vert \leqslant \vert \delta X_{\text{max}} \vert^2 e^{-\alpha_R t}
		\begin{cases}
			\frac{t^2}{2} & \beta_R = \gamma_R = 0 \\
			\frac{e^{\gamma_R t} -1}{\gamma_R^2} - \frac{t}{\gamma_R} & \beta_R = 0,  \; \gamma_R \neq 0  \\
			\frac{t e^{\beta_R t}}{\beta_R} - \frac{e^{\beta_R t} -1 }{\beta_R^2} & \beta_R \neq 0, \; \gamma_R = 0 \\
			\frac{e^{(\beta_R + \gamma_R)t} - 1 }{\gamma_R(\beta_R + \gamma_R) } - \frac{e^{\beta_R t} }{\beta_R \gamma_R} & \beta_R, \gamma_R \neq 0 
		\end{cases}
	\end{align}
where $x_R$ denotes the real part of the number $x \in \lbrace \alpha, \beta, \gamma \rbrace$. In the long-time limit one has for $\alpha_R > 0$,
	\begin{align}
	\label{eq:integral_limit_second_order_long_time}
		0 \leqslant I^{(2)}(\alpha, \beta , \gamma)  \leqslant \vert \delta X_{\text{max}} \vert^2 
		\begin{cases}
			0 & \beta_R = \gamma_R = 0 \\
			\frac{e^{(\gamma_R - \alpha_R) t} }{\gamma_R^2} & \beta_R = 0,  \; \gamma_R \neq 0  \\
			e^{(\beta_R- \alpha_R) t} \left( \frac{t }{\beta_R} - \frac{1 }{\beta_R^2} \right) & \beta_R \neq 0, \; \gamma_R = 0 \\
			\frac{e^{(\beta_R + \gamma_R- \alpha_R )t}  }{\gamma_R(\beta_R + \gamma_R) } - \frac{e^{(\beta_R - \alpha_R) t} }{\beta_R \gamma_R} & \beta_R, \gamma_R \neq 0 
		\end{cases}.
	\end{align}
The estimates~\eqref{eq:integral_limit_2} and~\eqref{eq:integral_limit_second_order_long_time} allow us to determine the terms in the expansion of the transmission amplitude $A$ that are non vanishing in the long-time limit.

\section{The limit $\Lambda \rightarrow 0$}
\label{appx:limit_lambda_zero}
In this Appendix we show that the transmission obtained in the case of a diagonalizable system matrix $L$ ($\Lambda \neq 0$, Sec.~\ref{sec:diagonalizable_L}) transitions smoothly into the transmission in the case of a non-diagonalizable system matrix $L$ ($\Lambda = 0$, Sec.~\ref{sec:non-diagonalizable_L}). By expanding the exponentials $\exp ([i\Delta + \Gamma \pm \Lambda ]t)$ in orders of $\Lambda t$, we find for the noise integrals in Eq.~\eqref{eq:A0_fp_fpq} of the main text,
	\begin{align}
	\label{eq:noise_integrals_powers_of_lambda}
	\begin{split}
		f_+ - f_- & = 2 \Lambda \lambda e^{- l t}\int_0^t dt_1 \delta X(t_1) e^{ l t_1} (t_1-t) + \mathcal{O} (\Lambda^2), \\
		\sum_{\pm} \pm \left( f_{\pm \pm} - f_{\mp \pm} \right) & = 4 \Lambda \lambda^2e^{-l t} \int_0^t dt_1 \delta X(t_1)  \int_0^{t_1} \delta X(t_2) e^{l t_2} (t_1-t) + \mathcal{O} (\Lambda^2), \\
		 \sum_{\pm} \left( f_{\pm \pm} - f_{\mp \pm} \right) & = 4 \Lambda^2 \lambda^2 e^{-l t} \int_0^t dt_1 \delta X(t_1)  \int_0^{t_1} \delta X(t_2) e^{l t_2} (t_1 t_2 + t_1 t - t t_2 - t_1^2) + \mathcal{O} (\Lambda^3),
	\end{split}
	\end{align}
where $l = i \Delta + \Gamma$. The powers of $\Lambda$ in these expressions cancel the negative powers of $\Lambda$ in the transmission amplitude~\eqref{eq:transmission_amplitude}, and in the limit $\Lambda \rightarrow 0 $ (this requires $\delta_0 = 0$) one finds
	\begin{align}
	\label{eq:A_limit_lambda_zero}
		\frac{A(t) }{A_0} = 1 - \frac{ \langle \sigma_z \rangle g^2}{ i \Delta + \gamma/2} \bigg[ i \tilde{I}_1(t) +  \tilde{I}_2(t) + \frac{\kappa - \gamma}{4} \tilde{I}_3(t)\bigg],
	\end{align}
where
	\begin{align}
	\label{eq:tilde_I}
	\begin{split}
		\tilde{I}_1(t) & = \lambda e^{- l t}\int_0^t dt_1 \delta X(t_1) e^{ l t_1} (t-t_1), \\
		\tilde{I}_2(t) & = \lambda^2e^{-l t} \int_0^t dt_1 \delta X(t_1)  \int_0^{t_1} \delta X(t_2) e^{l t_2} (t-t_1) , \\
		\tilde{I}_3(t) & =\lambda^2 e^{-l t} \int_0^t dt_1 \delta X(t_1)  \int_0^{t_1} \delta X(t_2) e^{l t_2} (t_1 t_2 + t_1 t - t_2 t - t_1^2).
	\end{split}
	\end{align}
Hence, the limit $\Lambda = 0$ exists. Moreover, one can easily see that Eq.~\eqref{eq:A_limit_lambda_zero} is precisely the transmission~\eqref{eq:transmission_critical_T} obtained in Sec.~\ref{sec:non-diagonalizable_L} if $\tilde{I}_1(t) =  I_1(t)$, $\tilde{I}_2(t) =  I_2(t)$ and $\tilde{I}_3(t) =  I_3(t)$ with $I_{1,2,3}$ as defined in Eq.~\eqref{eq:noise_integrals_critical}. The latter three equalities hold true for any fluctuations $\delta X(t)$ that are of exponential order in $t$ as can be shown by writing $\delta X(t)$ in its Laplace representation in Eqs.~\eqref{eq:noise_integrals_critical} and~\eqref{eq:tilde_I} and integrating the resulting expression in Eq.~\eqref{eq:tilde_I} by parts.

\section{Mean and variance of the transmission}
\label{appx:mean_and_variance}

\subsection{Perturbation series in the noise}

In general, the normalized transmission has the form of a perturbation series $\vert A \vert/\vert A_0 \vert = \sqrt{ 1 + \sum_{n=1}^{\infty} A_n(\delta X) }$, where the terms $A_n(\delta X)$ are of order $n$ in the noise $\delta X$ and $\vert A_0 \vert$ is the noise-free transmission. We find for the first and second non-central moments, respectively,
	\begin{align}
		\frac{\llangle \vert A \vert \rrangle}{\vert A_0\vert} & = 1 + \frac{1}{2} \llangle A_1 (\delta X) \rrangle +  \frac{1}{2}  \llangle A_2 (\delta X) \rrangle -  \frac{1}{8} \llangle A_1 (\delta X)^2 \rrangle + \mathcal{O}(\delta X^3), \\
		\frac{\llangle \vert A \vert^2 \rrangle}{\vert A_0\vert^2 }  & = 1 +  \llangle A_1 (\delta X) \rrangle +  \llangle A_2 (\delta X) \rrangle + \mathcal{O}(\delta X^3).
	\end{align}
Consequently, the variance reads
	\begin{align}
	\label{eq:general_variance}
		\frac{\text{Var} \left(\vert A \vert \right)}{\vert A_0 \vert^2} = \frac{ \llangle \vert A \vert^2 \rrangle - \llangle \vert A \vert \rrangle^2}{\vert A_0 \vert^2} = \frac{1}{4} \left( \llangle A_1(\delta X)^2 \rrangle -  \llangle A_1(\delta X) \rrangle^2 \right)  + \mathcal{O}(\delta X^3).
	\end{align}
The second term vanishes in the case of zero-mean noise, while the first term is non-zero in general. For the normalized transmission amplitude in Eq.~\eqref{eq:transmission_amplitude} of Sec.~\ref{sec:diagonalizable_L} one has 
	\begin{align}
		A_1(\delta X) = \text{Re} \left( \frac{ i  \langle \sigma_z \rangle g^2}{2 \Lambda \left( i \Delta_q + \gamma/2 \right)}   \left[ f_+(t) - f_-(t) \right] \right),
	\end{align}
and for the normalized transmission amplitude in Eq.~\eqref{eq:transmission_critical_T} of Sec.~\ref{sec:non-diagonalizable_L} we find
	\begin{align}
		A_1(\delta X) = -\text{Re} \left( \frac{ - \langle \sigma_z \rangle g^2}{  i \Delta + \gamma/2} I_1(t)\right).
	\end{align}
Since one has to square the real part of the noise integrals \textit{before} averaging when computing the mean or variance, it is not possible to rewrite these quantities in terms of convolutions. As a result, the figure of merit in Sec.~\ref{sec:perturbation_theory_in_noise} is chosen to be the second non-central moment, i.e., the averaged transmission probability, $\llangle \vert A \vert^2 \rrangle = \llangle \vert A \vert \rrangle^2 + \text{ Var} (\vert A \vert)$.

\subsection{Perturbation series in the qubit-photon coupling}

In the dispersive regime discussed in Sec.~\ref{sec:perturbation_theory_in_g} the transmission probability is a perturbation series in the small parameter $\varepsilon = g/\text{max} \lbrace \vert \delta_0 \vert, \vert \kappa - \gamma \vert \rbrace$, $\vert A \vert = \sum_{n=0}^{\infty} \varepsilon^n A_n$. As is shown in the supplementary material of Ref.~\cite{Mutter2022}, the variance reads
	\begin{align}
	\label{eq:variance}
		&\text{Var} (\vert A \vert ) =   \varepsilon^2 \left( \llangle A_1^2 \rrangle - \llangle A_1 \rrangle^2 \right) + \mathcal{O} \left( \varepsilon^3 \right)
	\end{align}
if $A_0$ is not affected by noise. This is true in Sec.~\ref{sec:perturbation_theory_in_g} where $A_0 \equiv A_{\infty} $ is the noise-free transmission through an empty cavity. In the long-time limit the term linear in $g$ vanishes, $A_1 \rightarrow 0$. As a consequence, the variance of $\vert A \vert $ vanishes up to quadratic order in $\varepsilon$ and hence $\llangle \vert A \vert^2 \rrangle = \llangle \vert A \vert \rrangle^2 + \mathcal{O} (\varepsilon^3)$.

\section{Explicit form of the functions $\varrho$ and $\psi_j$}
\label{appx:psi_functions}
Using the notation $\Lambda_1 = \text{Re} (\Lambda) $ and $\Lambda_2 = \text{Im} (\Lambda)$, the function $\varrho$ appearing in Eq.~\eqref{eq:transmission_convolutions} of the main text reads
	\begin{align}
	\begin{split}
	\label{eq:phi}
		&\varrho = \frac{4 \Lambda_2 \gamma + 8 \Lambda_1 \Delta_q}{\vert \Lambda \vert^2} \frac{2 \Lambda_2 (  \Lambda_1 \Lambda_2 - \Gamma \Delta) - \Lambda_1 (\Gamma^2 - \Delta^2 + \Lambda_2^2 - \Lambda_1^2)}{(\Gamma^2 - \Delta^2 + \Lambda_2^2 - \Lambda_1^2)^2 + 4 (\Gamma \Delta - \Lambda_1 \Lambda_2)^2} \\
		& \qquad + \frac{4 \Lambda_1 \gamma - 8 \Lambda_2 \Delta_q}{\vert \Lambda \vert^2} \frac{ \Lambda_2 (\Gamma^2 - \Delta^2 + \Lambda_2^2 - \Lambda_1^2) + 2 \Lambda_1 (  \Lambda_1 \Lambda_2 - \Gamma \Delta)}{(\Gamma^2 - \Delta^2 + \Lambda_2^2 - \Lambda_1^2)^2 + 4 (\Gamma \Delta - \Lambda_1 \Lambda_2)^2}.
	\end{split}
	\end{align}
Moreover, the set of functions $\lbrace \psi_j : j \in \lbrace  1,2,3,4,5 \rbrace \rbrace$ appearing in Eq.~\eqref{eq:transmission_convolutions} of the main text has elements of the form
	\begin{align}
		&\psi_1 = 4 \langle \sigma_z \rangle g^2, \quad\psi_{2/3} = \frac{\nu_{\pm} f_{\pm} + \mu_{\pm} g_{\pm}}{2}, \quad\psi_{4/5} = \frac{\nu_{\pm} h_{\pm} + \mu_{\pm} k_{\pm}}{2},
	\end{align}
where the positive (negative) sign belongs to the even (odd) index in the latter two equations,
	\begin{align}
	\begin{split}
			\nu_{\pm} & = \left[ \frac{\Lambda_1^2 - \Lambda_2^2}{\vert \Lambda \vert ^4} \frac{\gamma}{2} - 2 \Delta_q \frac{\Lambda_1 \Lambda_2}{\vert \Lambda \vert^4} \right] \left[ \gamma - \kappa \pm 4 \Lambda_1 \right] - \left[ \frac{\Lambda_1 \Lambda_2}{\vert \Lambda \vert^4} \gamma + \Delta_q \frac{\Lambda_1^2 - \Lambda_2^2}{\vert \Lambda \vert ^4} \right] \left[ 2 \delta_0 \mp 4 \Lambda_2 \right], \\
			\mu_{\pm} & = \left[ \frac{\Lambda_1^2 - \Lambda_2^2}{\vert \Lambda \vert ^4} \frac{\gamma}{2} - 2 \Delta_q \frac{\Lambda_1 \Lambda_2}{\vert \Lambda \vert^4} \right] \left[ 2 \delta_0 \mp 4 \Lambda_2  \right] + \left[ \frac{\Lambda_1 \Lambda_2}{\vert \Lambda \vert^4} \gamma + \Delta_q \frac{\Lambda_1^2 - \Lambda_2^2}{\vert \Lambda \vert ^4} \right] \left[ \gamma - \kappa \pm 4 \Lambda_1 \right], 
	\end{split}
	\end{align}	 
and
	\begin{align}
	\begin{split}
		f_{\pm} & = \frac{(\Gamma \pm \Lambda_1)^2}{(\Gamma \pm \Lambda_1)^2 + (\Delta \pm \Lambda_2)^2}	 - \frac{(\Gamma \pm \Lambda_1)(\Gamma \mp \Lambda_1)}{(\Gamma \mp \Lambda_1)^2 + (\Delta \mp \Lambda_2)^2}, \; \; \quad h_{\pm} = - \frac{\Delta \pm \Lambda_2 }{(\Gamma \pm \Lambda_1)^2 + (\Delta \pm \Lambda_2)^2}	 + \frac{\Delta \mp \Lambda_2 }{(\Gamma \mp \Lambda_1)^2 + (\Delta \mp \Lambda_2)^2}, \\
		g_{\pm} & = - \frac{(\Delta \pm \Lambda_2)(\Gamma \pm \Lambda_1) }{(\Gamma \pm \Lambda_1)^2 + (\Delta \pm \Lambda_2)^2}	 + \frac{(\Delta \mp \Lambda_2)(\Gamma \pm \Lambda_1) }{(\Gamma \mp \Lambda_1)^2 + (\Delta \mp \Lambda_2)^2}, \quad k_{\pm} = - \frac{\Gamma \pm \Lambda_1}{(\Gamma \pm \Lambda_1)^2 + (\Delta \pm \Lambda_2)^2}	 + \frac{\Gamma \mp \Lambda_1}{(\Gamma \mp \Lambda_1)^2 + (\Delta \mp \Lambda_2)^2}.
	\end{split}
	\end{align}

\section{Fourier transforms of the convolution kernels}
\label{appx:Fourier_transform_kernels}
In this Appendix we display the explicit expressions for the Fourier transforms of the convolution kernels in Eqs.~\eqref{eq:convolution_kernels} and \eqref{eq:convolution_kernels_critical} of the main text. These are needed to compute the noise power spectral density from the measured convolutions by the convolution theorem. Given a function $f(\Delta)$, its Fourier transform is computed according to the convention $\tilde{f} (\tau) = (1  / 2 \pi) \int_{-\infty}^{\infty}  f(\Delta) e^{-i \Delta \tau} d \Delta$.

\subsection{Diagonalizable system matrix}
We first consider the kernels in Eq.~\eqref{eq:convolution_kernels} of the main text. Since $\vert \Lambda_1 \vert < \Gamma$ the kernel $K_1$ has two first-order poles in the upper half plane at $ \pm \Lambda_2 +i (\Gamma \mp \Lambda_1)$ and two first-order poles in the lower half plane at $\pm \Lambda_2 - i (\Gamma \mp \Lambda_1)$. The kernels $K_2$ and $K_4$ have two first-order poles at $- \Lambda_2 \pm i(\Gamma + \Lambda_1)$, and the kernels $K_3$ and $K_5$ have two-first order poles at $ \Lambda_2 \pm i(\Gamma - \Lambda_1)$. Closing the contour in the upper (lower) half plane for $\tau<0$ ($\tau>0$), one finds by applying the residue theorem,
	\begin{align}
	\label{eq:Fourier_transform_kernels}
	\begin{split}
		\tilde{K}_1(\tau) & = \begin{cases}
			\frac{e^{-i \Lambda_2 \tau + (\Gamma - \Lambda_1) \tau }}{8(\Gamma - \Lambda_1)(\Lambda_1 + i \Lambda_2)(\Gamma - i \Lambda_2)} - \frac{e^{i \Lambda_2 \tau + (\Gamma + \Lambda_1) \tau }}{8(\Gamma + \Lambda_1)(\Lambda_1 + i \Lambda_2)(\Gamma + i \Lambda_2)} & \tau < 0 \\
			 \frac{e^{-i \Lambda_2 \tau - (\Gamma - \Lambda_1)\tau }}{8(\Gamma - \Lambda_1)(\Lambda_1 - i \Lambda_2)(\Gamma ü+ i \Lambda_2)} - \frac{e^{i \Lambda_2 \tau - (\Gamma + \Lambda_1) \tau}}{8(\Gamma + \Lambda_1)(\Lambda_1 - i \Lambda_2)(\Gamma - i \Lambda_2)} & \tau \geq 0 
		\end{cases} , \\
		\tilde{K}_{2/3} (\tau) & = \frac{e^{ \pm i \Lambda_2 \tau - (\Gamma \pm \Lambda_1) \vert \tau \vert }}{2(\Gamma \pm \Lambda_1) }, \quad  \tilde{K}_{4/5}(\tau) = - \frac{i}{2} \text{sgn} (\tau) e^{ \pm i \Lambda_2 \tau - (\Gamma \pm \Lambda_1) \vert \tau \vert },
	\end{split}
	\end{align}
where once again the positive (negative) sign is associated with the even (odd) index in the latter two equations. The expressions for $\tilde{K}_{4/5}(\tau) $ are discontinuous at $\tau = 0$ because the kernels $K_{4/5}(\Delta)$ are not absolutely integrable. However, the functions are square integrable, and so the Fourier transforms exist.

\subsection{Non-diagonalizable system matrix}
Each kernel in Eq.~\eqref{eq:convolution_kernels_critical} of the main text has two second order poles at $\Delta_{\pm} = \pm i \Gamma$. Closing the contour in the upper (lower) half plane for $\tau<0$ ($\tau>0$), one finds by again applying the residue theorem
	\begin{align}
	\begin{split}
	\label{eq:Fourier_transform_kernels_critical}
		&\tilde{K}_1(\tau) = \left( \frac{\vert \tau \vert }{ \Gamma^2} + \frac{1}{ \Gamma^3} \right)\frac{e^{-\Gamma \vert \tau \vert }}{4}, \quad \tilde{K}_2(\tau) = \left(  \frac{\kappa - \gamma}{4} \vert \tau \vert + 1  \right) e^{-\Gamma \vert\tau \vert }, \quad \tilde{K}_3(\tau) = -  i \left(  \frac{\kappa - \gamma}{4} \tau  + \frac{\text{sgn} (\tau)}{2} \right) e^{-  \Gamma \vert \tau \vert  }.
	\end{split}
	\end{align}
Note that - as was the case with the kernels $K_{4/5}$ when treating a diagonalizable system matrix above - the Fourier transform of $K_3(\Delta)$ is discontinuous at $\tau = 0$ because $K_3$ is not absolutely integrable but exists because the kernel is square integrable.

\section{Extracting the convolutions}
\label{appx:extract_convolutions_full}

In this Appendix we present specific schemes to extract the convolutions from the measured transmission. We treat the case where the analytical solution is obtained as a perturbation series in the noise in Sec.~\ref{appx:extract_convolutions} and the case where it is obtained as a perturbation series in the qubit-photon coupling in Sec.~\ref{appx:extract_convolutions_g}.

\subsection{Perturbation series in the noise}
\label{appx:extract_convolutions}

We first present a specific scheme to extract the convolutions from the measured transmission according to the results of Sec.~\ref{sec:perturbation_theory_in_noise} under the assumption of independent control over the parameters $\delta_0$ (via the qubit frequency $\omega_q$), $\omega_p$ and $g$ (e.g. via the detuning in a semiconductor charge qubit). Working at the point of equal decay rates, $\kappa = \gamma$, one has $\text{Re} (\Lambda) = 0$ and $\text{Im} (\Lambda) = \sqrt{(\delta_0/2)^2 -  \langle \sigma_z \rangle g^2}$. In the following we aim to change the parameters such that the sum $\sum_j \psi_j \mathcal{C}_j$ appearing in the expression for the averaged transmission probability can be recorded as a function of the detuning $\delta_0$ while the convolutions $\mathcal{C}_j$ are kept constant at any given $\Delta$.

The requirement that the $\mathcal{C}_j$ are unchanged at the distinct parameter configurations (i.e. at different $\delta_0$) can be achieved by keeping $\text{Im} (\Lambda)$ and $\Delta$ fixed as can be seen from Eq.~\eqref{eq:convolution_kernels} of the main text. $\text{Im} (\Lambda)$ ($\Delta$) can be kept constant by compensating the change in $\delta_0$ by a change in $g$ ($\omega_p$). Specifically, one has the dependences
	\begin{align}
	\begin{split}
		 g^2(\delta_0) =  \frac{4 \langle \sigma_z (\delta_{0,0}) \rangle g_0^2 + \delta_0^2 - \delta_{0,0}^2}{4\langle \sigma_z (\delta_0) \rangle} = \frac{4  \tanh([\omega_c - \delta_{0,0}]/2T ) g_0^2 + \delta_0^2 - \delta_{0,0}^2}{4 \tanh([\omega_c - \delta_0]/2T ) }  , \quad \omega_p(\delta_0) = \omega_{p,0} + \frac{\delta_{0,0} - \delta_0}{2},
	\end{split}
	\end{align}
where $\delta_{0,0}$, $g_0$ and $\omega_{p,0}$ are the initial parameter values.

After extracting the values of the convolution $\mathcal{C}_j$ at a given $\Delta$ by fitting the expression for the transmission probability in Eq.~\eqref{eq:transmission_convolutions} as a function of $\delta_0$ to the measured data, one may change $\Delta$ by changing $\omega_{p,0} \rightarrow \omega_{p,1}$ and repeat the procedure for as many points of $\Delta$ as required for computing the Fourier transform of one of the convolutions to the desired accuracy.

\subsection{Perturbation series in the coupling constant}
\label{appx:extract_convolutions_g}

Finally, we turn to the case discussed in Sec.~\ref{sec:perturbation_theory_in_g} of the main text in which the transmission is obtained as a perturbation series in the qubit-photon coupling $g$, and detail how the real and imaginary parts of the convolution in Eq.~\eqref{eq:second_derivative_2} can be extracted from the measured data. By varying both the probe frequency $\omega_p$ and the qubit frequency $\omega_q$ it is possible to keep $\Delta_q = \omega_q - \omega_p$ (and hence the convolution $\mathcal{C}$) constant while $\Delta_c = \omega_c - \omega_p$ is changed. Measuring the averaged transmission and thereby the real part of the ANI for two distinct values $\Delta_{c,1}$ and $\Delta_{c,2}$, one may obtain an inhomogeneous system of linear equations for the real and imaginary parts of the convolution $\mathcal{C}$,
	\begin{align}
	\label{eq:SLG}
		\begin{pmatrix}
		\alpha_1 (\Delta_{c,1}) & \alpha_2 (\Delta_{c,1}) \\
		\alpha_1 (\Delta_{c,2}) & \alpha_2 (\Delta_{c,2})
	\end{pmatrix}
	\begin{pmatrix}
	\text{Re} (\mathcal{C}) \\
	\text{Im} (\mathcal{C})
	\end{pmatrix}
	=
	\begin{pmatrix}
	D^2_{\mathcal{N}}(\Delta_{c,1}) \\
	D^2_{\mathcal{N}}(\Delta_{c,2}) 
	\end{pmatrix},
	\end{align}
where $D^2_{\mathcal{N}}(\Delta_{c}) = - d^2 \text{Re} \llangle \mathcal{N} \rrangle / d \lambda^2 \vert_{\lambda = 0}$ is the curvature of the real part of the ANI and
	\begin{align}
	\begin{split}
		\alpha_1 (\Delta_c)  =  \frac{\left[ \kappa \left( \gamma^2 - 4 \Delta_q^2 \right) - 8 \gamma \Delta_q \Delta_c \right] }{(\kappa^2 + 4 \Delta_c^2)(\gamma^2 + 4 \Delta_q^2)^2}, \quad	\alpha_2 (\Delta_c)  = \frac{ \left[ 2 \Delta_c \left( \gamma^2 - 4 \Delta_q^2 \right) - 4 \kappa \Delta_q \gamma \right] }{(\kappa^2 + 4 \Delta_c^2)(\gamma^2 + 4 \Delta_q^2)^2}.
	\end{split}
	\end{align}
The system of linear equations may be solved for the real and imaginary parts of the convolution if
	\begin{align}
	\det \begin{pmatrix}
		\alpha_1 (\Delta_{c,1}) & \alpha_2 (\Delta_{c,1}) \\
		\alpha_1 (\Delta_{c,2}) & \alpha_2 (\Delta_{c,2})
	\end{pmatrix}
	=\frac{\Delta_{c,1} -\Delta_{c,2} }{(\kappa^2 + 4\Delta_{c,1}^2) (\kappa^2 + 4 \Delta_{c,2}^2) (\gamma^2 + 4 \Delta_q^2)^4}  2 \kappa \left[ 16 \gamma^2 \Delta_q^2 - (\gamma^2 - 4 \Delta_q^2)^2 \right] \neq 0.
	\end{align}
Hence, a unique solution exists if $16 \gamma^2 \Delta_q^2 - (\gamma^2 - 4 \Delta_q^2)^2 \neq 0$, which is true for all but four values of $\Delta_q$,
	\begin{align}
	\label{eq:Delta_q_critical}
		\Delta_q^{p_1,p_2} =  \left( \frac{(-1)^{p_1}}{2} + \frac{(-1)^{p_1+p_2} }{\sqrt{2}} \right) \gamma , \quad p_1, p_2 \in \lbrace 0,1 \rbrace.
	\end{align}
Alternatively, one may use two distinct values of $\kappa$ instead of $\Delta_c$ to extract the convolutions. Due to the dependence of the functions $\alpha_{1,2}$ on the system parameters the same analysis holds, and as before the convolution $\mathcal{C}$ can be extracted for all detunings excluding the four values given in Eq.~\eqref{eq:Delta_q_critical}.

\section{Comparison of the two perturbation approaches}
\label{appx:comparison}

In this Appendix we show that the transmissions obtained in Secs.~\ref{sec:perturbation_theory_in_noise} and~\ref{sec:perturbation_theory_in_g} (denoted here by $A_{\delta X}$ and $A_g$, respectively) agree in the appropriate limit, i.e., 
	\begin{align}
	\label{eq:equality_claim}
		A_{\delta X}\left( \text{up to order } g^2 \right) = A_g \left( \text{up to order } \delta X^2 \right).
	\end{align}
We start by expanding the transmission $A_g$ in Eq.~\eqref{eq:long_time_transmission} in orders of $\lambda \delta X (t)$ and hence $\lambda \mathcal{X} (t_1,t_2) = \lambda \int_{t_1}^{t_2} \delta X(t) dt$, $\exp (-i \lambda \mathcal{X}(t_1,t_2)) = 1 - i \lambda \mathcal{X}(t_1,t_2) - \lambda^2 \mathcal{X}(t_1,t_2)^2/2 + ...$, yielding after evaluation of the integral featuring the zeroth-order term,
	\begin{align}
	\label{eq:expansion_A_g}
		A_g = A_{\infty} \left[ 1 + \langle \sigma_z \rangle g^2 \left\lbrace \frac{1}{qc} - i \lambda    I_{g}^{(1)} (t) - \frac{\lambda^2}{2} I_{g}^{(2)}(t) \right\rbrace\right] + \mathcal{O} \left(\delta X^3 \right),
	\end{align}
where $c = i \Delta_c + \kappa/2$, $q = i \Delta_q + \gamma/2$ and
	\begin{subequations}
	\begin{align}
	\label{eq:expansion_integrals_g}
		I_{g}^{(1)}(t) &= e^{-c t } \int_0^t dt_1 e^{  (c-q)t_1}  \int_0^{t_1} dt_2 e^{q t_2 } \mathcal{X} (t_1, t_2), \\
	\label{eq:expansion_integrals_g_2}
		I_{g}^{(2)}(t) & = e^{-c t } \int_0^t dt_1 e^{ (c - q) t_1}   \int_0^{t_1} dt_2 e^{q t_2 } \mathcal{X} (t_1, t_2)^2.
	\end{align}
	\end{subequations}
Similarly, we may expand the transmission $A_{\delta X}$ in Eq.~\eqref{eq:transmission_amplitude} in orders of $g$, again working up to second order. It is useful to first take the limit in the quantities $A_0$, $\Lambda$ and $l_{\pm}$, yielding
	\begin{align}
	\begin{split}
		A_0  = A_{\infty} \left( 1 +  \frac{\langle \sigma_z \rangle g^2}{qc} \right), \quad 	\Lambda  = \frac{c-q}{2} + \frac{ \langle \sigma_z \rangle g^2}{c-q}, \quad l_+  = c + \frac{ \langle \sigma_z \rangle g^2}{c-q}, \quad
		l_-  = q- \frac{ \langle \sigma_z \rangle g^2}{c-q}.
	\end{split}
	\end{align}
Inserting these expressions into Eq.~\eqref{eq:transmission_amplitude}, we find
	\begin{align}
	\label{eq:expansion_A_dX}
		A_{\delta X} = A_{\infty} \left[ 1 + \langle \sigma_z \rangle g^2 \left\lbrace \frac{1}{qc} - i \lambda I_{\delta X}^{(1)} (t) - \frac{\lambda^2}{2} I_{\delta X}^{(2)}(t)  \right\rbrace \right] + \mathcal{O} \left(g^3 \right),
	\end{align}
where
    \begin{subequations}
	\begin{align}
	\label{eq:expansion_integrals_dX}
		I_{\delta X}^{(1)}(t) &= \frac{1}{q(c-q)} \left( e^{-q t } \int_0^t  dt_1\delta X(t_1) e^{q t_1 }   -  e^{-c t } \int_0^t dt_1 \delta X(t_1) e^{c t_1 }  \right), \\
	\label{eq:expansion_integrals_dX_2}
		I_{\delta X}^{(2)}(t) & =  \frac{2}{q(c-q)} \left( e^{-q t} \int_0^t dt_1 \delta X(t_1)\int_0^{t_1} dt_2 \delta X(t_2) e^{q t_2 }   - e^{-c t } \int_0^t dt_1 \delta X(t_1) e^{ (c -  q) t_1 } \int_0^{t_1}dt_2 \delta X(t_2) e^{q t_2 }    \right).
	\end{align}
    \end{subequations}
Comparing Eqs.~\eqref{eq:expansion_A_g} and~\eqref{eq:expansion_A_dX}, it is clear that they agree at zeroth order in $\lambda$. Therefore, we must only show that $I_{g}^{(1)} = I_{\delta X}^{(1)}$ and $I_{g}^{(2)} = I_{\delta X}^{(2)}$ to proof the claim~\eqref{eq:equality_claim}. In the following we will show the equality at order $\lambda^1$, and the tools used in the proof can be applied to demonstrate the equality at order $\lambda^2$ in an analogous (yet more tedious) approach that we do not show explicitly.

We proceed by rewriting $I_{g}^{(1)}$ in Eq.~\eqref{eq:expansion_integrals_g} using $\mathcal{X}(t_1,t_2) = \mathcal{X}(t_2)  - \mathcal{X}(t_1) $ with $\mathcal{X} (t) = \int_0^t \delta X(s) ds$ and neglecting all terms that vanish in the long-time limit,
	\begin{align}
		I_{g}^{(1)} (t) = e^{-c t } \left( \frac{1}{q} \int_0^t dt_1 \mathcal{X} (t_1)e^{c t_1 }  - \int_0^t dt_1 e^{(c - q) t_1 } \int_0^{t_1} dt_2\mathcal{X}(t_2)   e^{q t_2}  \right).
	\end{align}
Integrating the $t_2$-integral in the second term by parts and using $\partial_t \mathcal{X} (t) = \delta X (t)$, we find
	\begin{align}
		I_{g}^{(1)} (t) = e^{-c t} \int_0^t dt_1 e^{ (c - q) t_1 } \int_0^{t_1} dt_2 \delta X(t_2) \frac{ e^{q t_2 } }{q} .
	\end{align}
Another partial integration in the $t_1$-integral then completes the proof at first order in $\lambda$,
	\begin{align}
	\begin{split}
		I_{g}^{(1)} (t)  = \frac{ e^{-c t } }{q(c-q)}  \left(  e^{(c - q) t }  \int_0^t dt_1 \delta X(t_1) e^{q t_1 }  -   \int_0^t dt_1 \delta X(t_1) e^{c t_1 }  \right) = I_{\delta X}^{(1)} (t).
	\end{split}
	\end{align}
	
The equality $I_{g}^{(2)} = I_{\delta X}^{(2)}$ at second order in $\lambda$ can be shown similarly by starting from Eq.~\eqref{eq:expansion_integrals_g_2}, expanding $\mathcal{X}(t_1,t_2)^2 = \mathcal{X}(t_1)^2 + \mathcal{X}(t_2)^2 - 2 \mathcal{X}(t_1)\mathcal{X}(t_2)$ and repeatedly integrating by parts to arrive at Eq.~\eqref{eq:expansion_integrals_dX_2}.
\twocolumngrid
\bibliography{charge_noise_cQED}
\end{document}